%% file: main.tex
\documentclass[journal,10pt]{IEEEtran}

\newcommand{\LRT}[2]{%
	\mathrel{\mathop\gtrless\limits^{#1}_{#2}}%
}
\newcommand\scalemath[2]{\scalebox{#1}{\mbox{\ensuremath{\displaystyle #2}}}}

\ifCLASSINFOpdf
\else
   \usepackage[dvips]{graphicx}
\fi
\usepackage{url}

\usepackage{graphicx}
\usepackage{cite}
\usepackage{amsmath,amssymb,amsfonts}
\usepackage{algorithmic}
\usepackage{textcomp}
\usepackage{amssymb}
\usepackage{amsmath}
\usepackage{graphicx}
\usepackage{graphics}
\usepackage{epstopdf}
\usepackage{caption}
\usepackage{enumitem}
\usepackage{xcolor,stfloats}
\usepackage{color}
\usepackage{float}
\usepackage{multicol}
\usepackage{mathtools}
\usepackage{array}
\usepackage{subfig}
\usepackage{setspace}
\usepackage{algorithm}
\usepackage{algorithmic}
\usepackage{bm}
\usepackage{cases}

\def\BibTeX{{\rm B\kern-.05em{\sc i\kern-.025em b}\kern-.08em
    T\kern-.1667em\lower.7ex\hbox{E}\kern-.125emX}}

\newlength\myindent
\DeclareMathOperator*{\argmax}{arg\,max}
\DeclareMathOperator*{\argmin}{arg\,min}

\allowdisplaybreaks[4]
\begin{document}

\title{Adaptive Radar Detection and Bearing Estimation in the Presence of Unknown Mutual Coupling}

\author{Augusto Aubry,~\IEEEmembership{Senior Member,~IEEE}, Antonio De Maio,~\IEEEmembership{Fellow,~IEEE}, Lan~Lan,~\IEEEmembership{Member,~IEEE}, and Massimo Rosamilia,~\IEEEmembership{Student Member,~IEEE}

\thanks{A. Aubry, A. De Maio, and M. Rosamilia are with the Department of Electrical Engineering and Information Technology, University of Naples ``Federico II'', I-80125 Naples, Italy, and also with the National Inter-University Consortium for Telecommunications, 43124 Parma, Italy (e-mail: augusto.aubry@unina.it; ademaio@unina.it; massimo.rosamilia@unina.it). (Corresponding author: Antonio De Maio.).}
\thanks{{Lan Lan is with the National Key Laboratory of Radar Signal Processing, Xidian University, Xi'an 710071, China, (e-mail: lanlan$@$xidian.edu.cn);}}
\thanks{The work of Lan Lan was supported in part by the National Nature Science Foundation of China (Nos. 62101402, 61931016, 62071344), Young Elite Scientists Sponsorship Program by CAST (2021QNRC001), and China Postdoctoral Science Foundation (Nos. 2021TQ0261, 2021M702547).}
}

\markboth{}%
{Shell \MakeLowercase{\textit{et al.}}: Bare Demo of IEEEtran.cls for Journals}
\maketitle

\begin{abstract}
This paper deals with joint adaptive radar detection and target bearing estimation in the presence of mutual coupling among the array elements. First of all, a suitable model of the signal received by the multichannel radar is developed via a linearization procedure of the Uniform Linear Array (ULA) manifold around the nominal array looking direction together with the use of symmetric Toeplitz structured matrices to represent the mutual coupling effects.
Hence, the Generalized Likelihood Ratio Test (GLRT) detector is evaluated under the assumption of homogeneous radar environment. Its computation leverages a specific Minorization-Maximization (MM) framework, with proven convergence properties, to optimize the concentrated likelihood function under the target presence hypothesis.
Besides, when the number of active mutual coupling coefficients is unknown, a Multifamily Likelihood Ratio Test (MFLRT) approach is invoked. During the analysis phase, the performance of the new detectors is compared with benchmarks as well as with counterparts available in the open literature which neglect the mutual coupling phenomenon. The results indicate that it is necessary to consider judiciously the coupling effect since the design phase, to guarantee performance levels close to the benchmark.
\end{abstract}


\section{Introduction}
Target detection is a long-standing key task in standard radar/sonar applications~\cite{Pomr1, SonarBook, DeMaioBook, liu2022multichannel}. {It has been the} subject of {plenty} of articles in the open literature, mainly devoted to the development of adaptive detectors {(as well as to their analysis)} capable of operating in the presence of undesired disturbance, hostile interference, and clutter.~\cite{SonarBook,Kelly2,liu2022multichannel,381910,414779, AMFKelly,543869,839972,4203039, Pomr1,DeMaioBook,7493946}.
Generally, to accomplish the detection task, at the design stage the received signal (under the assumption of target presence) is deemed as the superposition of the {target} echo and the interference-plus-noise contribution, which is usually {modeled as the} realization of a zero-mean Gaussian process with an unknown and possibly structured covariance matrix. Besides, the standard homogeneous radar environment assumption is invoked, where a set of secondary (training) data, free of useful contributions from the target, enables the estimation of the unknown interference covariance matrix and the derivation of adaptive architectures~\cite{Kelly2,AMFKelly,DeMaioBook, liu2022multichannel,414779,543869,839972}.
Under the mentioned circumstances, the target detection problem is formulated in terms of a binary statistical hypothesis test, whose optimal solution (in the Neyman-Pearson sense) is given by the Likelihood Ratio Test (LRT)~\cite{vantrees4, DeMaioBook, KayDetection, Pomr1}. However, it demands perfect knowledge of the likelihood functions under the two hypotheses including their parameters tied up to either the target characteristics or the interference covariance matrix. In practical situations, such parameters are unknown and demand an estimation procedure. This pushes toward the development of implementable receivers based on sub-optimal approaches, such as for instance the Generalized Likelihood Ratio (GLR), where the unknowns are replaced by their Maximum-Likelihood (ML) estimates~\cite{van2001detection, KayDetection}. 

Once the presence of the target is established in the Cell Under Test (CUT), the estimation process of the target bearing could be accomplished by means of monopulse~\cite{Nickel}, generalized monopulse~\cite{gen-monopulse} or other bespoke techniques, {construing} detection and estimation {as} two different signal processing tasks~\cite{YANG2018509,9468638}.
However, it is worth pointing out that in the open literature some architectures implementing jointly detection-estimation have been successfully derived, to reveal the target presence and simultaneously provide its accurate angular bearing state~\cite{Lan_GLRT, AubryTSP}. 
The successful achievement of the detection/estimation {processes} requires handling some challenges. Among them, the angular uncertainty of the received signal with respect to (w.r.t.) the pointing direction as well as the presence of mutual coupling effects within the array, both introducing mismatches between the actual and the presumed steering vector{~\cite{besson,4133018, 1561887, 5604325, 4203039, 9082601}}. As to the former, it can be accounted for at the design level by linearizing the array manifold around the look-direction and modeling the target steering vector as the superposition of the pointing direction signature plus another term {due to} the actual angle offset w.r.t. the nominal array looking direction~\cite{AubryTSP}.
{As to the latter, in phased arrays the fields radiated by one antenna can be received by the other elements, engendering the phenomenon of mutual coupling. This involves an alteration of the electromagnetic characteristics of each array element caused by leakage phenomena from the radiating elements in its vicinity. Mutual coupling is affected by a plurality of factors, including number, type, and relative orientation of each antenna element, as well as inter-element distance~\cite{coupling_review}}. That said, the presence of mutual coupling among the radiating elements could affect the radar performance, degrading radar resolution capability, robustness to interference of adaptive algorithms, and target Direction of Arrival (DOA) estimation accuracy~\cite{coupling_review, 8123841}.
In the open literature, several studies address the effects of mutual coupling on radar detection~\cite{Landi, Fiorini}, sidelobe blanking~\cite{Fiorini2}, and adaptive beamforming~\cite{6189025, 7945283, iet-spr_20070198}. Furthermore, \cite{FrielPascal2} {is} focused on the impact of coupling on the performance of Space-Time Adaptive Processing (STAP) techniques, whereas~\cite{FrielPascal3} refers to wideband DOA estimation. Several {references dealing} with narrowband target DOA estimation problem {in} the presence of mutual coupling can be found in~\cite{Friedlander-TAP,142628, 1321335, 4805279,7880661,8661532}.

Nevertheless, the problem of jointly detecting the target and estimating its bearing while accounting for mutual coupling and the DOA uncertainty has not yet been addressed in the open literature. 
Aiming at filling this gap, a simultaneous detection and target bearing estimation procedure, developed for a Uniform Linear Array (ULA) affected by mutual coupling, is proposed in this paper.
To accomplish the two tasks simultaneously, both the unknown DOA displacement (w.r.t. the looking direction) and the coupling phenomenon are {suitably} modeled at the design phase, namely, the actual steering vector is expressed as the product of a mutual coupling matrix and an approximated steering vector depending linearly on DOA displacement.
As to the mutual coupling matrix, it can be well described by a symmetric banded Toeplitz matrix leveraging the assumption that the mutual coupling coefficients are inversely proportional to the distance between elements and may be neglected for sufficiently spaced antennas~\cite{Friedlander-TAP, LiaoMutualC, Wu-coupling, 29832}. The identifiability of the unknown parameters for the developed signal model {is also investigated}.

Then, the target detection problem is formulated assuming a homogeneous {radar} interference environment and addressed resorting to the Generalized LRT (GLRT)~\cite{KayDetection,DeMaioBook} and the Multifamily LRT (MFLRT)~\cite{kay2005multifamily} strategies. The former requires perfect knowledge of the number of mutual coupling coefficients, while the latter can be framed as a generalization of the GLRT which incorporates the unknown model order inference. The derived architectures demand for the ML estimation of both the coupling coefficients and the target DOA displacement, which are computed by means of an {ad-hoc} iterative procedure based on the Minorization-Maximization (MM) framework. 
The convergence properties of the devised procedure are also formally proven.
{In addition,} the Constant False Alarm Rate (CFAR) behavior of the proposed decision strategies is investigated, proving that they ensure the bounded CFAR property.
{Last but not least,} the extension of the methods to include a second processing stage, leveraging an additional linearization of the array manifold around the current bearing estimate, is also presented.

During the analysis phase, the performance of the proposed adaptive architectures is assessed in terms of Probability of Detection ($P_d$) and Root Mean Square Error (RMSE) of target bearing. As to the detection capabilities, benchmark and standard receivers, are included for comparison purposes. The estimation performance is compared to the Cram\'{e}r-Rao Bound (CRB), computed for both the actual and the linearized array model.
The numerical results highlight the potentialities of the proposed architectures {to realize} both the detection and the estimation task simultaneously, corroborating the capabilities of the devised strategies to counteract steering vector mismatches induced by {the} mutual coupling {phenomenon}. Besides, the two-stage processing schemes show a general improvement of detection and estimation performance {as} compared {with} the single stage counterparts.

The paper is organized as follows. 
The signal model accounting for the presence of mutual coupling and target DOA uncertainty is given in Section II.
The design of the joint detection-estimation architectures {and their relevant properties are} addressed in Section III. Section IV deals with the computation of the CRB for both the actual and the linearized array manifold case. 
The detection and estimation performance of the {proposed techniques} is analyzed in Section IV, whereas conclusions and future research avenues are discussed in Section VI.

\subsection{Notation}
Boldface is used for vectors $\bm{a}$ (lower case), and matrices $\bm{A}$ (upper case). The $(k, l)$-entry (or $l$-entry) of a generic matrix $\bm{A}$ (or vector $\bm{a}$) is indicated as $\bm{A}(k, l)$ (or $\bm{a}(l)$). $\bm{I}$ and ${\bm{0}}$ denote respectively the identity matrix and the matrix with zero entries (their size is determined from the context). The transpose and the conjugate transpose operators are denoted by the symbols $(\cdot)^{\mathrm{T}}$ and $(\cdot)^\dagger$, respectively. The trace of the matrix ${\bm{A}} \in \mathbb{C}^{N\times N}$ is indicated with $\operatorname{tr}\{\bm{A}\}$.
$\mathbb{R}^N$ and ${\mathbb{C}}^N$ are respectively the sets of $N$-dimensional column vectors of real and complex numbers. ${\mathbb{H}}^{N}_{++}$ represents the set of $N\times N$ Hermitian positive definite matrices. $\mathbb{T}^{BS}_P$ represents the set of banded symmetric Toeplitz matrices of order P.
The letter $j$ represents the imaginary unit (i.e., $j=\sqrt{-1}$). For any complex number $x$, $|x|$ indicates the modulus of $x$. Moreover, for any $\bm{x} \in \mathbb{C}^N$, $\|\bm{x}\|$ denotes the Euclidean norm. 
Let $f(\bm{x}, \bm{y}) \in \mathbb{R}$ be a real-valued function, $\nabla_{\bm{x}} f(\bm{x}, \bm{y})$ denotes the gradient of $f(\cdot)$ w.r.t. $\bm{x}$, with the partial derivatives arranged in a column vector.

\section{Signal Model}
Let us consider a {monostatic radar equipped with an antenna} array that collects spatial data using a narrow-band {ULA} composed of $N$ {elements}. {After} down-conversion, pulse compression, and fast-time sampling, the echo signal from a prospective target at distance $R$ and azimuth $\theta_0$, {with respect} to the array boresight, is given by
{\begin{equation}\label{eq:echo_sign}
		a\bm{p}(u_0),
\end{equation}
where $a$ is an unknown complex parameter embedding target backscatter and channel propagation effects, $u_0$ denotes the angular position of the target in the space of directional cosine, i.e., $u_0=\sin({\theta_0})$, and $\bm{p}(u_0)$ indicates the spatial steering vector $\bm{p}(u)$ evaluated at $u_0$.
Specifically,
\begin{equation}\label{eq:standard_steering_vector}
	\bm{p}(u) = [1, e^{j\frac{2\pi}{\lambda_0}d u}, \dots, e^{j\frac{2\pi}{\lambda_0}(N-1)d u}]^\mathrm{T} \in \mathbb{C}^N,
\end{equation}
where $\lambda_0$ is the radar operating wavelength and $d$ is the inter-element spacing (typically set as $d=\lambda_0/2$).

Following the same approach as in~\cite{AubryTSP}, the steering vector of the received echo signal can be approximated via the Linearized Array Manifold (LAM) at the nominal array looking direction $\bar{u}$, with a resulting functional dependency of the array manifold on the directional cosine offset $\Delta u=u_0-\bar{u}$, namely
\begin{equation}
	{\bm{p}(u) \approx\;} \bm{p}_a(\Delta u) = \bm{p}(\bar{u}) + \frac{\partial \bm{p}(\bar{u})}{\partial u} \Delta u = \bm{p} + \dot{\bm{p}}_u\Delta u 
\end{equation} 
with $\bm{p}=\bm{p}(\bar{u})$ and $\dot{\bm{p}}_u=\frac{\partial \bm{p}(\bar{u})}{\partial u}$, respectively.

So far, an ideal steering vector has been considered. However, in practice, the actual steering vector experiences mutual coupling, which could {lead to} some mismatches between the ideal array manifold considered at the design stage and the actual one{~\cite{4203039}}. 
To address this issue, the coupling effects must {be} {accounted for} at the signal processor design level, which entails modeling the actual steering vector as~\cite{LiaoMutualC}
\begin{equation}\label{eq:s_v_mutual_c}
	\bm{p}_m(u) = \bm{C}\bm{p}(u) {\in \mathbb{C}^N},
\end{equation} 
where
\begin{equation}\label{eq:matrixC}
	\begin{aligned}
		\bm{C}={\left[ \scalemath{0.8}{\begin{array}{ccccccccc}
			1 & c_1 & \cdots & c_{P-1} & 0& \cdots& \cdots& \cdots&0 \\
			c_1 & 1 & c_1 & \cdots & c_{P-1} &0 &\cdots &\cdots &0 \\
			\vdots & \ddots & \ddots & \ddots & \ddots & \ddots & \ddots& \ddots&\vdots \\
			c_{P-1} & \cdots & c_1 & 1 & c_1 & \cdots & c_{P-1} &\cdots & 0\\
			0& \ddots & \ddots & \ddots & \ddots & \ddots & \ddots & \ddots & \vdots \\
			\vdots& \ddots& c_{P-1} & \cdots & c_1 & 1 & c_1 & \cdots & c_{P-1} \\
			\vdots& \ddots& \ddots& \ddots & \ddots & \ddots & \ddots & \ddots & \vdots \\
			0& \cdots&\cdots & 0& c_{P-1} & \cdots & c_1 & 1 & c_1 \\
			0& \cdots& \cdots& \cdots& 0& c_{P-1} & \cdots & c_1 & 1
		\end{array}}\right]}
	\end{aligned}
\end{equation}
 represents the {$N \times N$} banded symmetric Toeplitz matrix of mutual coupling~\cite{Friedlander-TAP, LiaoMutualC, 29832}, with $\mathrm{c}_{i} {\in \mathbb{C}}, i=1,\dots, P-1$, the $i$-th mutual coupling coefficient. Model~\eqref{eq:s_v_mutual_c} is supported by empirical and physical considerations. In fact, the coupling effects between two sensors reduce as their distance increases, and can be practically considered negligible for sensors whose {separation} is large enough, i.e., according to~\eqref{eq:matrixC}, $P$ times the inter-element spacing~\cite{Friedlander-TAP, LiaoMutualC}.
 With reference to a ULA, in Fig.~\ref{fig:1}, a pictorial representation of the mutual coupling effects between the $i$-th antenna and its $2(P-1)$ nearest array elements (assumed present) is illustrated. In particular, the different colors reflect the symmetries in the electromagnetic field leakage. Before proceeding further, let us consider as case study a ULA composed of $N=16$ elements with $P=3$, $c_1=0.7$ and $c_2 = 0.4$. The cosine similarity between the ideal and the actual steering vector, i.e.,
\begin{equation}\label{eq:cos_mis}
		\mathrm{cos}_s (u) = \frac{{| \bm{p}_m(u)^\dagger \bm{p}(u) |} }
		{ \|\bm{p}_m(u)\|     \| \bm{p}(u)\| } ,
\end{equation}
{versus $u$} is reported in Fig.~\ref{fig:2}.
Inspection of the figure reveals that for some $u$, corresponding {approximately} to $\theta \in [-60^\circ,-30^\circ] \cup [30^\circ,60^\circ]$, the mismatch {induced} by the mutual coupling is {considerable}, with values {of} $\mathrm{cos}_s{(u)} < 0.8$. In {this circumstance}, the performance of standard signal processing architecture {could} degrade severely. {Hence, it becomes mandatory} the development of robust adaptive strategies accounting, at the design stage, for the possible presence of {mutual} coupling between the array elements.
To further investigate the effects of the {mutual coupling} when the nominal receive direction lies in one of the previously {mentioned} {angular regions}, let us consider the cosine similarity between the actual steering vector {at} $u_0 = \sin(35^\circ)$ and the ideal one, i.e., 
\begin{equation}\label{eq:cos_mis_2}
	\mathrm{cos}_s(u; u_0) = \frac{{| \bm{p}_m(u_0)^\dagger \bm{p}(u) |} }
	{ \|\bm{p}_m(u_0)\|     \| \bm{p}(u)\| },
\end{equation}
computed for several values of the directional {actual target DOA} cosine $u$. The result is reported in Fig.~\ref{fig:3} assuming the same scenario as in Fig.~\ref{fig:2}. The curve highlights that there is a displacement of $-1.38^\circ$ between the peak angle of the cosine similarity and the true DOA, which pinpoints the influence of mutual coupling on the DOA estimation process if it is not properly {modeled} in the processing architecture. It is also worth mentioning that, in general, the coefficients {$c_i$ and their number}, i.e., $P-1$, {might} not be known at the design level.

Now, considering the linearization of the steering vector in~\eqref{eq:s_v_mutual_c} around the radar pointing direction in conjunction with the coupling effects, \eqref{eq:s_v_mutual_c} can be approximated as
\begin{equation}\label{eq:actual_mc_sv}
	\bm{p}_{am}(\Delta u) = \bm{C}\bm{p}_a(\Delta u) =  \bm{C}\bm{p} +  \bm{C} \dot{\bm{p}}_u\Delta u .
\end{equation}

This implies that the useful echo signal~\eqref{eq:echo_sign} can be written as
\begin{equation} \label{eq:B_pa}
\begin{aligned}
	a \bm{p}_{am}(\Delta u) &= a\bm{C}\bm{p}_a(\Delta u) = \bm{B}\bm{p}_a(\Delta u) \\&= \bm{H}(\Delta u) \bm{b} \in \mathbb{C}^N ,
\end{aligned}
\end{equation}
where
	\begin{equation}
		\bm{B}=  a\bm{C} =  b_0 \bm{I} + \sum_{m=1}^{P-1} b_m \bm{D}_m \in \mathbb{C}^{N \times N},
	\end{equation}
	with {$\bm{D}_m$ the $N\times N$ matrix having 1s on its $m$-th upper and lower diagonals, and zeros elsewhere},
	\begin{eqnarray}
	&&	\bm{H}(\Delta u) = (\tilde{\bm{D}} + \Delta u \dot{\bm{D}}),\\
	&& \label{eq:D_tilde} \tilde{\bm{D}} = [\bm{p}, \bm{D}_1 \bm{p}, \dots, \bm{D}_{P-1} \bm{p}] \in \mathbb{C}^{N \times P},	\\
	&&	 \label{eq:D_point} \dot{\bm{D}} = [\dot{\bm{p}}_u, \bm{D}_1 \dot{\bm{p}}_u, \dots, \bm{D}_{P-1} \dot{\bm{p}}_u]  \in \mathbb{C}^{N \times P}, \\
	&&	\bm{b} = {[a, a\,c_1, \dots, a \,c_{P-1}]^{\mathrm{T}}} \in \mathbb{C}^{P}.
	\end{eqnarray}

\begin{figure}[t]
	\centering
	\includegraphics[width=0.95\linewidth]{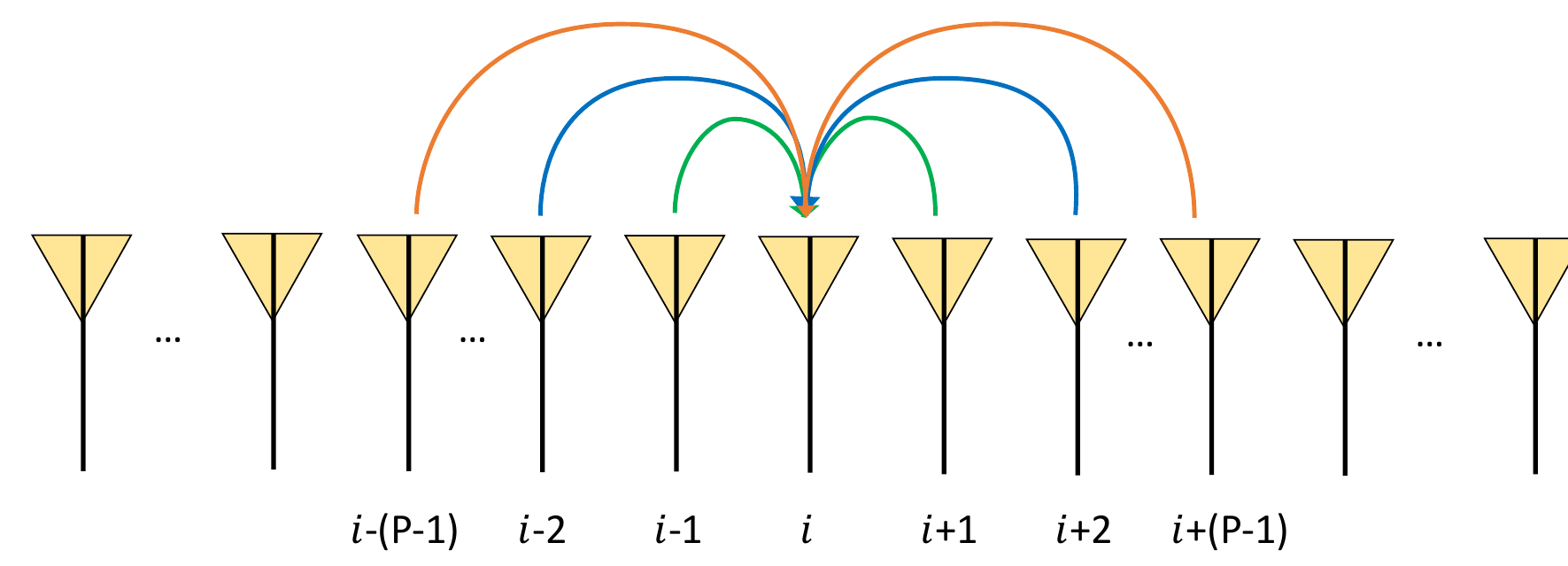}
	\caption{A notional representation of the mutual coupling {effects} between the $i$-th antenna and its {$2(P-1)$} nearest array elements {(assumed present)}.}\label{fig:1}
\end{figure}
\begin{figure}[t]
	\centering
	\includegraphics[width=0.99\linewidth]{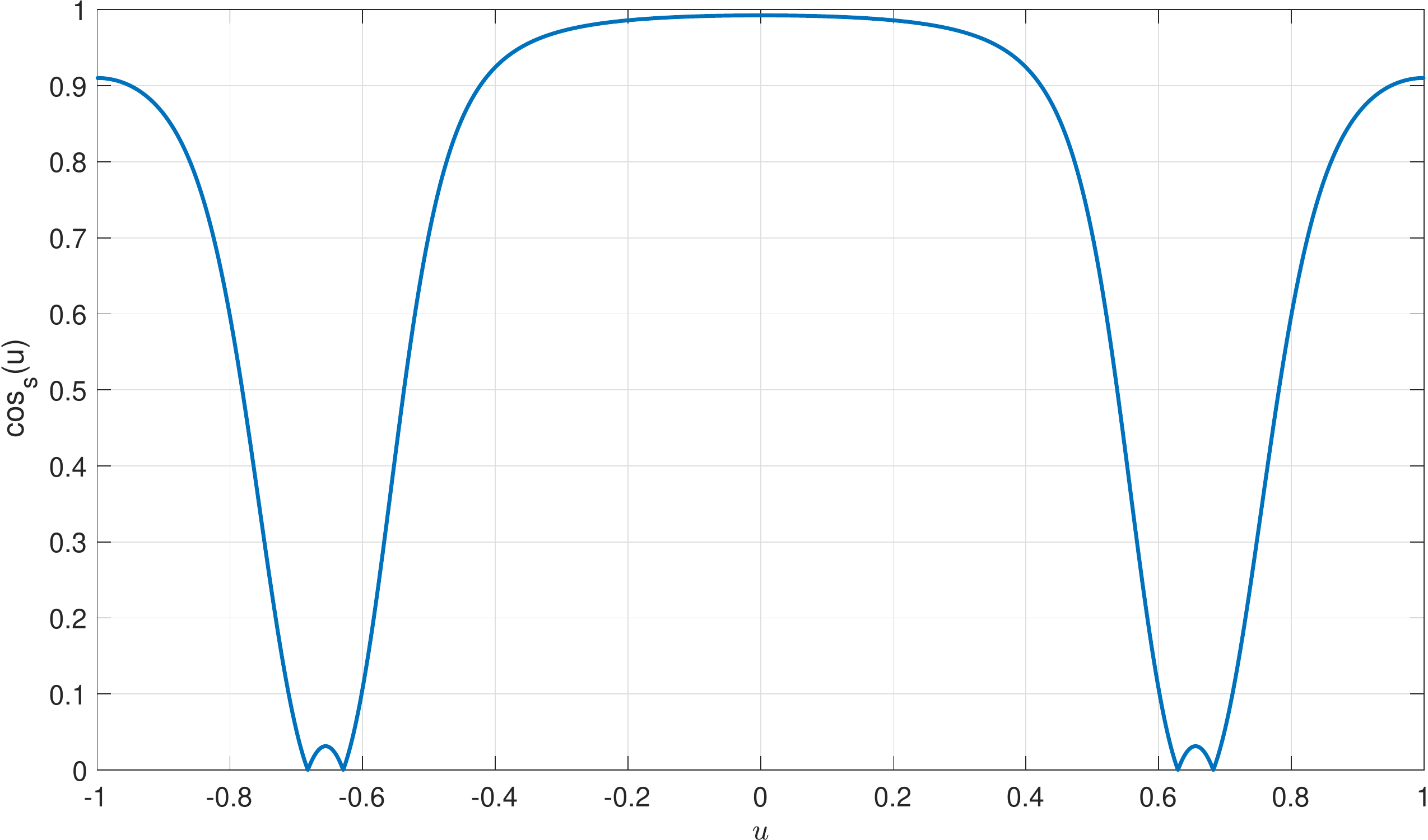}
	\caption{Cosine similarity{~\eqref{eq:cos_mis}} between the ideal and the actual steering vectors {vs $u$. A ULA with $N=16$ elements, $P=3$, $c_1=0.7$, and $c_2 = 0.4$, is considered}.}\label{fig:2}
\end{figure}
\begin{figure}[t]
	\centering
	\includegraphics[width=0.99\linewidth]{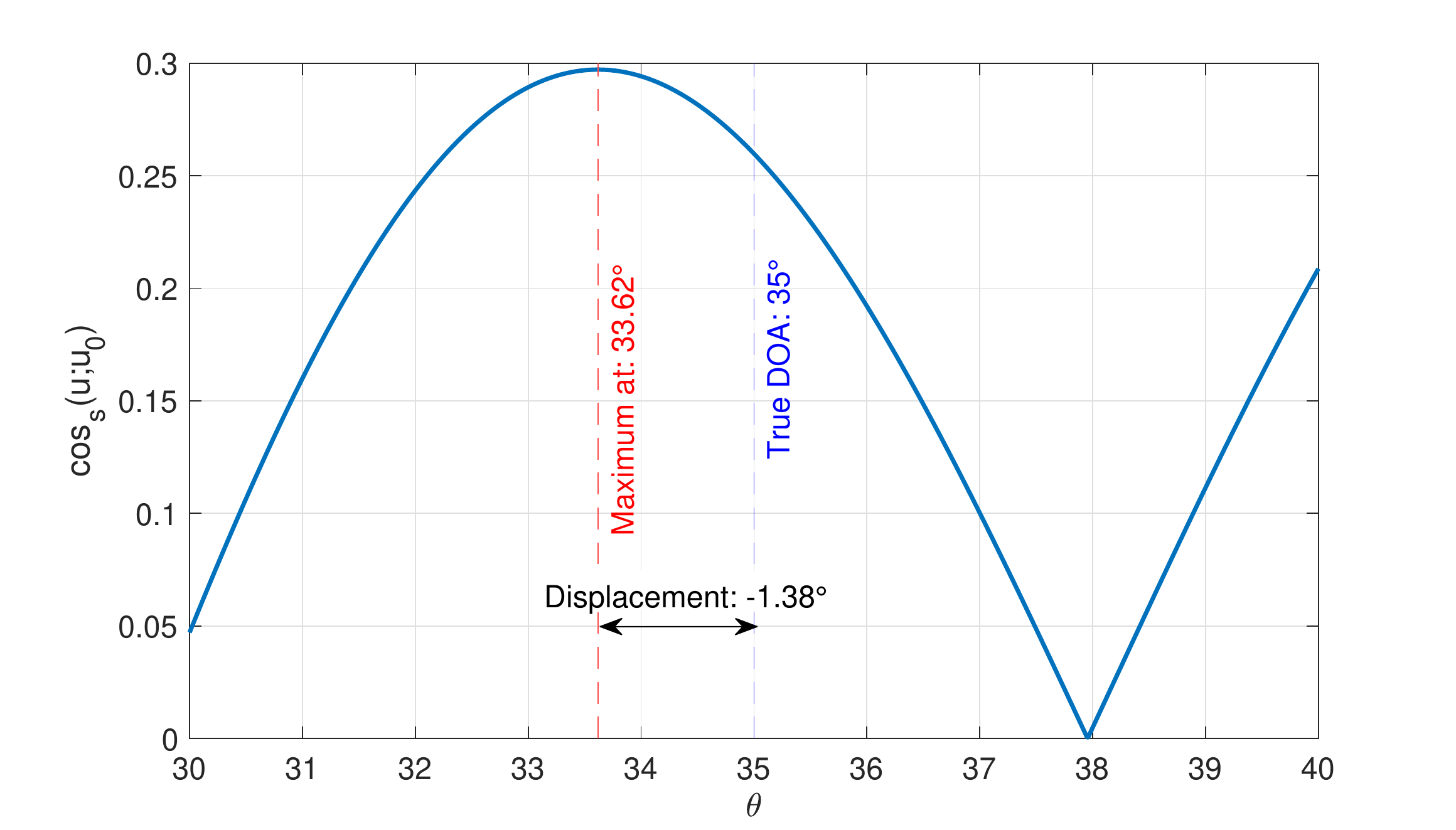}
	\caption{Cosine similarity~{\eqref{eq:cos_mis_2}} between {$p(u)|_{u=\sin(\theta)}$ and $p_m(u_0)$ vs $\theta$, assuming $u_0 = \sin(35^\circ)$. A ULA with $N=16$ elements, $P=3$, $c_1=0.7$, and $c_2 = 0.4$, is considered. The true {DOA} and the peak {angle} of the cosine similarity are {highlighted} as {the} blue and red vertical lines, respectively}.}\label{fig:3}
\end{figure}

\subsection{Model Identifiability} Let us analyze the identifiability of the unknown parameters in the signal model~\eqref{eq:B_pa} by considering the equation
\begin{equation}\label{eq:identif1}
	 \bm{H}(\Delta u )\bm{b} = \bm{H}(\Delta u^\star )\bm{b}^\star ,
\end{equation}
with $\Delta u^\star$ and ${\bm{b}^\star}$ being the true values {of the parameters}. 
To claim model identifiability, \eqref{eq:identif1} must admit the only solution $\Delta u = \Delta u^\star$, $\bm{b} = \bm{b}^\star$.
In this respect, a sufficient condition for {solution} uniqueness is that $P\le N/2$ and $\bm{H}_1 = [\tilde{\bm{D}}, \dot{\bm{D}}]$ {is} full rank. In fact, let us observe that \eqref{eq:identif1} is equivalent to
\begin{equation} \label{eq:identif2}
	\left\{\begin{matrix*}[l]
			\bm{H}_1 \bm{b}_1=\bm{H}_1 \bm{b}_1^\star\\ 
		\bm{b}_1 = [\bm{b}^\mathrm{T}, \Delta u \bm{b}^\mathrm{T}]^\mathrm{T}
	\end{matrix*}\right. ,
\end{equation}
where $\bm{b}_1^\star = [{\bm{b}^\star}^\mathrm{T}, \Delta u^\star {\bm{b}^\star}^\mathrm{T}]^\mathrm{T}$.
{The first} equation \eqref{eq:identif2} {can be cast as}
\begin{equation}
	\bm{H}_1 (\bm{b}_1 - \bm{b}_1^\star) = \bm{0} ,
\end{equation}
which is a homogeneous system of equations {admitting} as unique (due to the full rank {assumption on} $\bm{H}_1$) solution the trivial one, i.e., $\bm{b}_1 = \bm{b}_1^\star$, which is also feasible to~\eqref{eq:identif2}.
Based on the above considerations, in the following it is supposed that $\bm{H}_1$ is full column rank.

\section{Target Detection Problem}
Assuming that the radar operates in a standard homogeneous environment~\cite{Kelly2, AMFKelly, AubryTSP, 5575453,4203039,1561887,839972,liu2022multichannel} {(and references therein)}, which allows for the collection of a set of $K$ secondary data (free of any useful target signal) with the same interference plus noise covariance matrix as the primary data, the binary hypothesis testing problem, pertaining to the target presence{/absence} within the Cell Under Test (CUT), can be formulated as 
\begin{equation}\label{eq:linearized_hypothesis_test}
	\left\{
	\begin{aligned}
		&\mathcal{H}_0 : \left\{
		\begin{aligned}
			&\bm{r} = \bm{n}\\
			&\bm{r}_k = \bm{n}_k \quad k=1,\dots, K
		\end{aligned}
		\right.\\
		&\mathcal{H}_1 : \left\{
		\begin{aligned}
			&\bm{r} = \bm{H}(\Delta u) \bm{b} + \bm{n} \\
			&\bm{r}_k = \bm{n}_k \quad k=1,\dots, K
		\end{aligned}
		\right.
	\end{aligned}
	\right.  ,
\end{equation}
where 
\begin{itemize}
\item $\bm{r}$ and $\bm{r}_k, k=1,\dots, K,$ denote the primary and the secondary data {vectors}, respectively;
\item $\bm{H}(\Delta u)$ is function of the unknown target DOA displacement w.r.t. the array pointing direction;
\item  $\bm{b}$ is the unknown vector {accounting for} both the complex received target echo return $a$ and the $P-1$ complex mutual coupling coefficients {$\mathrm{c}_{m}${'s}};
\item $\bm{n}$ and $\bm{n}_k$, $k=1,\dots,K$, are the interference plus noise components of the received snapshots, modeled as statistically independent, complex, zero-mean, circularly symmetric Gaussian random vectors with unknown positive definite covariance matrix
\begin{equation}\label{eq:homog_environment}
	\bm{M}=E[\bm{n}\bm{n}^\dagger]=E[\bm{n}_k\bm{n}_k^\dagger] \in {\mathbb{H}}^{N}_{++}, \quad k=1,\dots,K .
\end{equation}
\end{itemize}

The standard strategy based on the Neyman-Pearson criterion can be used to determine the best decision statistic for the hypothesis-testing problem~\eqref{eq:linearized_hypothesis_test}, that is, to obtain a detector able to maximizing the $P_d$ for a desired Probability of False Alarm ($P_{fa}$). Unfortunately, the resulting {decision rule} requires {the} perfect knowledge of {the unknowns} in the PDFs under both {the} hypotheses, which is clearly {unavailable} in real application contexts. {In other words}, there is no Uniformly Most Powerful (UMP) test for this problem. Consequently, it is necessary to design practically implementable receivers using suboptimal criteria, such as the GLRT, which {leverages} the ML estimation of {the unknown} parameters under both hypotheses.

Note that the number of unknowns {connected with} the array coupling coefficients, {namely} $P-1$, can be either assumed known, i.e., it can be periodically measured exploiting calibration procedures or predicted by means of some electromagnetic considerations, or {modeled as an additional} unknown {parameter}.
Therefore, in the following the GLRT is first developed assuming {that} the number of coupling coefficients {is} {known} at the design stage. Then, the case of {unknown} model order is considered and a decision rule is derived by means of the MFLRT criterion~\cite{kay2005multifamily, kay2}.

\subsection{Decision Statistic for Known Model Order}\label{section:known_model_order}
Assuming $P$ known, the target detection problem~\eqref{eq:linearized_hypothesis_test} can be {handled} resorting to the GLRT criterion
\begin{equation}\label{eq:constrained_GLRT}
	\frac{ \max\limits_{\bm{M} \in {\mathbb{H}}^{N}_{++},\bm{B}\in\mathbb{T}^{BS}_P,|\Delta u|\leq \alpha} f_{\mathcal{H}_1}(\bm{r}, \bm{r}_1, \dots, \bm{r}_k|\bm{M},a, \bm{B}, \Delta u)}
	{\max\limits_{\bm{M} \in {\mathbb{H}}^{N}_{++}}f_{\mathcal{H}_0}(\bm{r}, \bm{r}_1, \dots, \bm{r}_k|\bm{M})} \LRT{\mathcal{H}_1}{\mathcal{H}_0} \gamma  ,
\end{equation}
where $\gamma$ is the detection threshold set to ensure a desired $P_{fa}$,
\begin{equation} \label{eq:likelih_h0}
	f_{\mathcal{H}_0}(\bm{r}, \bm{r}_1, \dots, \bm{r}_k|\bm{M}) = \left[\frac{1}{\pi^N |\bm{M}|} e^{-\operatorname{tr}\{ \bm{M}^{-1} \bm{T}_0 \} } \right]^{K+1}
\end{equation}
and
\begin{equation} \label{eq:likelih_h1}
\begin{aligned}
	&f_{\mathcal{H}_1}(\bm{r}, \bm{r}_1, \dots, \bm{r}_k|\bm{M},\bm{b}, \Delta u) =\\ & \qquad \qquad \qquad \qquad \left[\frac{1}{\pi^N |\bm{M}|} e^{-\operatorname{tr}\{\bm{M}^{-1} \bm{T}_1 \}}\right]^{K+1}
\end{aligned}
\end{equation}
represent the likelihood function of the observations under the {$\mathcal{H}_0$ and $\mathcal{H}_1$ hypothesis, respectively}, with
\begin{equation}
	\bm{T}_0 = \frac{1}{K+1} \left( \bm{r} \bm{r}^\dagger + \sum_{k=1}^{K} \bm{r}_k \bm{r}_k^\dagger \right)
\end{equation}
and
\begin{equation} \label{eq:T1}
\begin{aligned}
	\bm{T}_1 = \frac{1}{K+1} \left( (\bm{r}-\bm{H}(\Delta u) \bm{b}) (\bm{r}-\bm{H}(\Delta u) \bm{b})^\dagger + \sum_{k=1}^{K} \bm{r}_k \bm{r}_k^\dagger \right).
\end{aligned}
\end{equation}

{Let us now describe the procedure necessary to perform the optimizations at the numerator and the denominator of~\eqref{eq:constrained_GLRT}}.
\subsubsection{Optimization w.r.t. $\bm{M}$}
{Based on} standard argumentation~\cite{Kelly2}, concentrating the likelihood functions~\eqref{eq:likelih_h0} and~\eqref{eq:likelih_h1} over $\bm{M}$ and taking their logarithm, the decision statistic in \eqref{eq:constrained_GLRT} is equivalent to
\begin{equation}\label{eq:glrt_stat_raw1}
\begin{aligned}
	l_G & = 2 (K+1)\log \left(\frac{|\bm{T}_0|}{\min\limits_{ \bm{b}, \;|\Delta u|\leq \alpha} |\bm{T}_1|}\right) \\ &= 2(K+1) \frac{1+\bm{r}_w ^\dagger\bm{r}_w}{1+\min\limits_{\bm{b}, \;|\Delta u|\leq \alpha} \left\| \bm{r}_w - \bm{H}_w(\Delta u) \bm{b}\right\|^2},
\end{aligned} 
\end{equation}
where $\bm{H}_w(\Delta u) = \bm{S}^{-1/2} \bm{H}(\Delta u)$ and $\bm{r}_w = \bm{S}^{-1/2} \bm{r}$ are the {quasi-}whitened {counterparts of $\bm{H}(\Delta u)$ and $\bm{r}$ respectively}, with $\bm{S} = \sum_{k=1}^{K} \bm{r}_k \bm{r}_k^\dagger$.

\subsubsection{Optimization w.r.t. $\bm{b}$}
The optimal solution $\hat{\bm{b}}$ {in~\eqref{eq:glrt_stat_raw1}} is given by\footnote{Notice that the assumption of $\bm{H}_1$ being full rank implies that{, for any $\Delta u$,} $\bm{H}(\Delta u)$ is full rank as well. As an immediate proof, assuming by contradiction $\bm{H}(\Delta u)$ be {not always full-rank}, i.e., {there exists a $\Delta u$ for which} at least one of its columns is a linear combination of the others, then one of the column of $\bm{H}_1$ becomes a linear combination of the others, which contradicts the hypothesis of $\bm{H}_1$ {being} full rank.}
\begin{equation}\label{eq:LS_b}
\begin{aligned}
		\hat{\bm{b}} = & \argmin_{\bm{b}} \left\| \bm{r}_w - \bm{H}_w(\Delta u) \bm{b}\right\|^2=  \bm{H}_w^o(\Delta u) \bm{r}_w ,
\end{aligned}
\end{equation}
where
\begin{equation}
	\bm{H}_w^o(\Delta u) = \left(\bm{H}_w^\dagger(\Delta u) \bm{H}_w(\Delta u) \right)^{-1} \bm{H}_w^\dagger(\Delta u)
\end{equation}
is the Moore-Penrose inverse of $\bm{H}_w(\Delta u)$.

Thus, concentrating~\eqref{eq:glrt_stat_raw1} over $\bm{b}$ {and ignoring irrelevant constants} yields
{\begin{equation}\label{eq:GLRT_1}
	\begin{aligned}
		\tau_{GLRT-LAM}= \frac{\max\limits_{|\Delta u|\leq \alpha}  \bm{r}_w^\dagger \bm{P}_H(\Delta u) \bm{r}_w}{1+\|\bm{r}_w\|^2}
	\end{aligned} ,
\end{equation}
where $\bm{P}_H(\Delta u) = \bm{H}_w(\Delta u) \bm{H}_w^o(\Delta u) $ is the projector onto $\bm{H}_w(\Delta u)$.}

\subsubsection{Optimization w.r.t. $\Delta u$}
Given the {decision statistic}~\eqref{eq:GLRT_1}, it is crystal clear that the heart of the problem is the derivation of a solution to the constrained optimization problem {at the numerator}, i.e., solving
\begin{equation} \label{eq:optimization_prob}
	\widehat{\Delta u} =  \argmax\limits_{|\Delta u|\leq \alpha} \bm{r}_w^\dagger \bm{P}_H(\Delta u) \bm{r}_w .
\end{equation}
Unfortunately, the objective function in~\eqref{eq:optimization_prob} is non concave and a closed-form solution cannot be derived. Besides, an {accurate} exhaustive discrete line search would entail a high computational complexity which {could} not be compliant with the timeline of a typical radar processor. Note also that the optimal solution can be conceptually derived nulling the first order derivative of the objective function in~\eqref{eq:optimization_prob}. Now, since this latter can be cast as the ratio of two polynomials, the optimal solution can be basically obtained finding the roots of a polynomial. However, both the evaluation of the polynomial coefficients and (more important) the execution of the root finding procedure are computationally expensive. Besides, the latter may suffer of numerical instabilities thus affecting the overall strategy reliability. 
	
In order to account for the aforementioned issues, in the following, the optimization problem~\eqref{eq:optimization_prob} is tackled {resorting to} the MM framework~\cite{Wu-MM, Ortega, Heiser}. In a nutshell, MM method is an iterative procedure, used to {tackle} {a} {challenging} {optimization problem} in an efficient and scalable way~\cite{palomarMM}. Specifically, each iteration of the method is composed of two steps. The former involves the computation of an appropriate {tight} minorant (surrogate function)~\cite{palomarMM}, which approximates the objective function around the {optimized solution derived} at the previous iteration. In the latter, the minorant is optimized and an updated {optimized} point is obtained.

Before proceeding further, let us rewrite the objective function in~\eqref{eq:optimization_prob} in terms of the auxiliary variables $\bm{y} = \bm{H}_w^\dagger(\Delta u) \bm{r}_w$ and $\bm{A} =  \bm{H}_w^\dagger(\Delta u) \bm{H}_w(\Delta u) {\in {\mathbb{H}}^{N}_{++}}$ as
\begin{equation}\label{eq:convex}
	{f(\Delta u)= \bm{y}^\dagger \bm{A}^{-1} \bm{y}\bigg|_{\bm{y}=\bm{H}_w^\dagger(\Delta u) \bm{r}_w,\atop \bm{A}=\bm{H}_w^\dagger(\Delta u) \bm{H}_w(\Delta u)}} .
\end{equation}
{As a first step of the MM framework, it is necessary to find a minorant to the objective function $f(\Delta u)$. In this respect, let us start from the functional extension of the Right-Hand Side (RHS) of~\eqref{eq:convex} considering it as an unrestricted function of $\bm{y} \in \mathbb{C}^N$ and $\bm{A} \in \mathbb{H}^{N}_{++}$, i.e.,
	\begin{equation}
		f(\bm{y}, \bm{A}) = \bm{y}^\dagger \bm{A}^{-1} \bm{y},
	\end{equation}
	which is jointly convex w.r.t. $\bm{y}$ and $\bm{A}$.
	Given a point {(}$\bm{y}_0$, $\bm{A}_0{)}$, and computing the tangent plane $f_a(\bm{y}, \bm{A}|\bm{y}_0, \bm{A}_0)$ to $f(\bm{y}, \bm{A})$ in $(\bm{y}_0, \bm{A}_0)$, the following inequality holds true
	\begin{equation}
		f(\bm{y}, \bm{A}) \ge f_a(\bm{y}, \bm{A}|\bm{y}_0, \bm{A}_0),
	\end{equation}
	where
	\begin{equation}\label{eq:tangent_plane}
	\begin{aligned}
			f_a(\bm{y}, \bm{A}|\bm{y}_0, \bm{A}_0) &= \bm{y}_0^\dagger \bm{A}_0^{-1} \bm{y}_0 \\ &+ 2 \operatorname{Re}\{ \nabla_{\bm{y}} f^\dagger(\bm{y}_0, \bm{A}_0) (\bm{y}-\bm{y}_0) \} 
		\\ & + \operatorname{tr}\{\nabla_{\bm{A}} f(\bm{y}_0, \bm{A}_0) (\bm{A} - \bm{A}_0)\}
	\end{aligned}
	\end{equation}
	with
	\begin{equation}
		\nabla_{\bm{A}} f(\bm{y}, \bm{A}) = -{\bm{A}^{-1}} \bm{y} \bm{y}^\dagger {\bm{A}^{-1}}
	\end{equation}
	and
	\begin{equation}
		\nabla_{\bm{y}} f(\bm{y}, \bm{A})  = \bm{A}^{-1} \bm{y} 
	\end{equation}
	denote the gradient of $f(\bm{y}, \bm{A}) $ w.r.t. $\bm{A}$ and $\bm{y}$, respectively.
	Now choosing $\bm{y}_0{^{(h-1)}}=\bm{H}_w^\dagger(\Delta u^{*(h-1)}) \bm{r}_w$ and $\bm{A}_0{^{(h-1)}}=\bm{H}_w^\dagger(\Delta u^{*(h-1)}) \bm{H}_w(\Delta u^{*(h-1)})$, with $\Delta u^{*(h-1)}$ the output of the MM algorithm at the $(h-1)-$th iteration, yields
	\begin{equation}\label{eq:minorization}
\begin{aligned}
			&	f(\Delta u) \ge \\ & {f_a(\bm{y},\bm{A}|{\bm{y}_0{^{(h-1)}},\bm{A}_0{^{(h-1)}}}){\bigg|_{{\bm{y}_0{^{(h-1)}}=\bm{H}_w^\dagger(\Delta u^{*(h-1)}) \bm{r}_w,\atop \bm{A}_0{^{(h-1)}}=\bm{H}_w^\dagger(\Delta u^{*(h-1)}) \bm{H}_w(\Delta u^{*(h-1)})}}}}\\&=f_a(\Delta u|\Delta u^{*(h-1)}),
\end{aligned}
	\end{equation}
	with equality if $\Delta u=\Delta u^{*(h-1)}$.}

{As per the second step of the MM algorithm}, {it} demands, at the $h$-th iteration, {the maximization of the RHS of~\eqref{eq:minorization}, namely} {(after some algebra)} the solution to
\begin{equation}\label{eq:opt_conv}
		\Delta u^{*(h)} = \argmax\limits_{|\Delta u|\leq \alpha} \rho |\Delta u|^2 + \zeta \Delta u + \gamma ,
\end{equation}
where
\begin{equation}\label{eq:rho}
	\rho = \operatorname{tr}\{	\nabla_{\bm{A}} f(\bm{y}_0{^{(h-1)}}, \bm{A}_0{^{(h-1)}}) \dot{\bm{D}}_w^\dagger \dot{\bm{D}}_w\},
\end{equation}
\begin{equation}\label{eq:zeta}
	\begin{aligned}
		\zeta = & 2 \operatorname{Re}\{ \nabla_{\bm{y}} f^\dagger(\bm{y}_0{^{(h-1)}}, \bm{A}_0{^{(h-1)}})  \dot{\bm{D}}_w^\dagger \bm{r}_w  \} \\& + \operatorname{tr}\{	\nabla_{\bm{A}} f(\bm{y}_0{^{(h-1)}}, \bm{A}_0{^{(h-1)}}) (\dot{\bm{D}}_w^\dagger \tilde{\bm{D}}_w + \tilde{\bm{D}}_w^\dagger \dot{\bm{D}}_w ) \}
	\end{aligned},
\end{equation}
and $\gamma$ {is} a constant value {functionally} independent of $\Delta u$, with $ \dot{\bm{D}}_w = \bm{S}^{-1/2} \dot{\bm{D}}$ and $ \tilde{\bm{D}}_w = \bm{S}^{-1/2} \tilde{\bm{D}}$.

It is now worth noting that, since {$\rho < 0$}, the objective function in~\eqref{eq:opt_conv} is strictly concave in $\Delta u$; therefore the optimal solution is given either by the global optimum for the unconstrained version of~\eqref{eq:opt_conv}}, i.e.,
\begin{equation}\label{eq:delta_u_star}
	\widetilde{\Delta {u}} = -\zeta/(2\,\rho),
\end{equation}
{if this solution is feasible}, i.e., $ |\widetilde{\Delta {u}}| \le \alpha$, or by the boundary point, {i.e., either $\alpha$ or $-\alpha$}, which maximizes~\eqref{eq:opt_conv}.
{To summarize, at the $h$-th iteration, the derived MM-based procedure yields the following estimate
\begin{equation}\label{eq:last_MM}
\Delta u^{*(h)} = \max(\min(\widetilde{\Delta {u}}, \alpha), -\alpha).
\end{equation}}

{Observe that {Problem~{\eqref{eq:optimization_prob}}} satisfies the following conditions:}
{
	\begin{enumerate}
		\item[C.1)] the feasible set {$\mathbb{S} = [-\alpha, \alpha]$} is closed and convex;
		\item[C.2)] $f_a(\Delta u_0|\Delta u_0) = f(\Delta u_0), \;\forall \Delta u_0 \in \mathbb{S}$;
		\item[C.3)] $f_a(\Delta u|\Delta u_0) \le f(\Delta u), \;\forall (\Delta u, \Delta u_0)  \in {\mathbb{S}^2}$;
		\item[C.4)] $f_a(\Delta u|\Delta u_0)$ is continuous in $(\Delta u,\Delta u_0), \forall (\Delta u, \Delta u_0)  \in {\mathbb{S}^2}$;
		\item[C.5)] {$f'_a(\Delta u|\Delta u_0)|_{\Delta u = \Delta u_0} = f'(\Delta u)|_{\Delta u = \Delta u_0}, \; \forall \Delta u_0 \in \mathbb{S}$}.
\end{enumerate}

{As consequence, due to} \cite[Theorem 1]{razaviyayn2013unified}, any limit point of the iterates generated by {the MM algorithm} is a stationary point of Problem~\eqref{eq:optimization_prob}.

In conclusion, the above procedure, {terminating with the computation of~\eqref{eq:last_MM},} is iteratively repeated until the exit condition $|f(\bm{y}_0{^{(h)}}, \bm{A}_0{^{(h)}}) - f(\bm{y}_0{^{(h-1)}}, \bm{A}_0{^{(h-1)}})| < \varepsilon$ is satisfied, with $\epsilon > 0$ {a} user-defined exit threshold.

A summary of the procedure is reported in \textbf{Algorithm~\ref{alg:1}}, where the method is initialized with ${{\Delta {u}}^\star}^{(0)} = 0$.
Therefore, denoting by $\widehat{\Delta u}_{\rm{LAM}}$ the output of \textbf{Algorithm~\ref{alg:1}}, the expression of the devised GLRT decision statistic  is given by
\begin{equation}\label{eq:GLRT}
	\tau_{GLRT-LAM} = \frac{1+\bm{r}_w ^\dagger\bm{r}_w}{1+ \bm{r}_w ^\dagger\bm{r}_w - \bm{r}_w^\dagger \bm{P}_H\left(\widehat{\Delta u}_{\rm{LAM}}\right)  \bm{r}_w } .
\end{equation}

\begin{algorithm}[tp]
	\caption{{Angle displacement estimation via MM}.}
	\label{alg:1}
	\textbf{Input:} ${\bm{r}}, {\bm{S}}, \bar{u}, \alpha, {N}, P, {\tilde{\bm{D}}, \dot{\bm{D}}},  \varepsilon$.\\
	\textbf{Output:} $\widehat{\Delta u}_{\rm{LAM}}$. 
	\begin{enumerate}
		\item[1.] Compute $\bm{r}_w = \bm{S}^{-1/2} \bm{r}$, $\tilde{\bm{D}}_w = \bm{S}^{-1/2} \tilde{\bm{D}}$, and $\dot{\bm{D}}_w = \bm{S}^{-1/2} \dot{\bm{D}}$;
		\item[2.] Set $h = 0$, ${{\Delta {u}}^\star}^{(h)} = 0$
		\item[3.] \textbf{repeat}
		\item[4.] \quad $h = h + 1$;
		\item[5.] \quad Compute $\bm{H}_{w0} = (\tilde{\bm{D}}_w + \Delta u^{*(h-1)} \dot{\bm{D}}_w)$;
		\item[6.] \quad Compute $\bm{y}_0{^{(h-1)}}=\bm{H}_{w0}^\dagger \bm{r}_w$ and $\bm{A}_0{^{(h-1)}}= \bm{H}_{w0}^\dagger \bm{H}_{w0}$;
		\item[7.] \quad Find ${{\Delta {u}}^\star}^{(h)}$ using~{\eqref{eq:last_MM}};
		\item[8.] \textbf{until} $|f(\bm{y}_0{^{(h)}}, \bm{A}_0{^{(h)}}) - f(\bm{y}_0{^{(h-1)}}, \bm{A}_0{^{(h-1)}})| < \varepsilon$.
		\item[9.] {Output $\widehat{\Delta u}_{\rm{LAM}} = {{\Delta {u}}^\star}^{(h)}$.}
	\end{enumerate}
\end{algorithm}

\subsubsection{Bounded CFARness}

It is now worth observing that the derived GLRT decision statistic {ensures the bounded CFAR property. To prove this claim, let us start from the scaled version of~\eqref{eq:glrt_stat_raw1}}, i.e.,
\begin{equation}
	\tau_{GLRT-LAM} = \frac{1+\bm{r}_w ^\dagger\bm{r}_w}{1+\min\limits_{\bm{b}_1 = [\bm{b}^\mathrm{T}, \Delta u \bm{b}^\mathrm{T}]^\mathrm{T}, \;|\Delta u|\leq \alpha} \left\| \bm{r}_w - \bm{H}_{1,w} \bm{b}_1\right\|^2},
\end{equation}
{and notice that}
\begin{equation} \label{eq:CFAR_GLRT}
	\begin{aligned}
		\tau_{GLRT-LAM} \le & \frac{1+\bm{r}_w ^\dagger\bm{r}_w}{1+\min\limits_{\bm{b}_1 \in \mathbb{C}^{2P}} \left\| \bm{r}_w - \bm{H}_{1,w} \bm{b}_1\right\|^2} \\ = & \frac{1+\bm{r}_w ^\dagger \bm{r}_w}{1+ \bm{r}_w^\dagger \bm{r}_w - \bm{r}_w^\dagger \bm{P}_{H1,w} \bm{r}_w} = \tau_{P},
	\end{aligned}
\end{equation}
where $\tau_{P}$ is a CFAR statistic~\cite{Kraut}, with $\bm{H}_{1,w} = \bm{S}^{-1/2} \bm{H}_{1}$ and $\bm{P}_{H1,w} = \bm{H}_{1,w} \left(\bm{H}_{1,w}^\dagger \bm{H}_{1,w} \right)^{-1} \bm{H}_{1,w}^\dagger$ {the projector onto the range span of $\bm{H}_{1,w}$}.

\subsection{Decision Statistics for Unknown Model Order}
In some situations, the number of significant mutual coupling coefficients $P-1$ can often be unknown both at the design and at the operative stage. In such {a} case, the target detection problem can be framed as a multiple composite alternative hypothesis testing problem, where each alternative hypothesis $\mathcal{H}_i, \; i=1,\, \dots, \bar{N}$, pertains to a different number of unknown signal parameters, i.e.,
\begin{equation}\label{eq:hyp_test_unknown_P}
	\left\{
	\begin{aligned}
		&\mathcal{H}_0 : \left\{
		\begin{aligned}
			&\bm{r} = \bm{n}\\
			&\bm{r}_k = \bm{n}_k \quad k=1,\dots, K
		\end{aligned}
		\right.\\
		&\mathcal{H}_i : \left\{
		\begin{aligned}
			&\bm{r} = \bm{B}_i\bm{p}_a(\Delta u) + \bm{n} \\
			&\bm{r}_k = \bm{n}_k \quad k=1,\dots, K
		\end{aligned}
		\right. ,\quad i=1,\dots, \bar{N}
	\end{aligned} ,
	\right. 
\end{equation}
with $\bar{N} \le N/2$ the {maximum}\footnote{Although from a mathematical point of view it should be considered $\bar{N} = N$, in general it is reasonable (according to physical or empirical considerations) to restrict {the range of values for} $\bar{N}$. Moreover, $\bar{N} \le N/2$ ensures model identifiability.} {allowed model order} and
\begin{equation}\label{eq:B_i}
	\bm{B}_i = b_0 \bm{I} + \sum_{m=1}^{i-1} b_m \bm{D}_m .
\end{equation} 
Remarkably, since the considered alternative hypotheses are nested, i.e., $\mathcal{H}_i \subset \mathcal{H}_j, \; i<j$, {the decision problem connected with}~\eqref{eq:hyp_test_unknown_P} can be tackled resorting to the MFLRT framework~\cite{kay2005multifamily}.
Thus, the target presence can be established according to {the decision rule}
\begin{equation}\label{eq_MFLRT}
\begin{aligned}
		&\tau_{MFLRT-LAM} = && \\ &\qquad \max\limits_{1\le i \le \bar{N}} \left\{ \left[l_G^{(i)} - (2i+1) \left(\log\left(\frac{l_G^{(i)}}{2i+1}\right)+1\right)\right] \right. &&\\ &\qquad \left. \;\; u\left(\frac{l_G^{(i)}}{ 2i+1 }-1\right)\right\} > \bar{\gamma} ,
\end{aligned}
\end{equation}
where {${2i+1}$ is the number of unknown parameters under the ${\mathcal{H}_i}$ hypothesis, which are related to the useful component of the received signal, i.e., the DOA displacement $\Delta u$ and the $i$ complex {mutual coupling coefficients}, $l_G^{(i)}$ denotes the GLRT statistic~\eqref{eq:glrt_stat_raw1} derived assuming $P=i$, $\bar{\gamma}$ is the threshold guaranteeing the demanded $P_{fa}$, and $u(t)$ is the {unit} step function, i.e., $u(t) = 1$ as long as $t \ge 0$ and zero elsewhere. Specifically, denoting by $\widehat{\Delta u}^{(i)}_{\rm{LAM}}$ the estimate of the offset obtained with \textbf{Algorithm~\ref{alg:1}} assuming $P=i$,
\begin{equation}
l_G^{(i)} =	2(K+1) \frac{1+\bm{r}_w ^\dagger\bm{r}_w}{1+  \|\bm{r}_w\|^2 - \bm{r}_w^\dagger \bm{P}_H\left(\widehat{\Delta u}^{(i)}_{\rm{LAM}}\right)  \bm{r}_w }.
\end{equation}

Let us now investigate the bounded CFARness of~\eqref{eq_MFLRT}. To this end, let us preliminary observe that the transformation involved in~\eqref{eq_MFLRT}, i.e.,
\begin{equation}
	\begin{aligned}
		g_i(x) =& \left[x - (2i+1) \left(\log \left(\frac{x}{2i+1}\right)+1\right)\right] \\ & u\left(\frac{x}{2i+1}-1\right),
	\end{aligned}
\end{equation}
which is function of $i \in \mathbb{N}$ and $x > 0$, exhibits two properties~\cite{kay2005multifamily}:
\begin{itemize}
	\item for any $i$, {$g_i(x)$} monotonically increases with $x$,
	\item $g_l(x) {\le} g_k(x)$ for any $l>k$, with $l,k \le \bar{N}$ and any $x>0$. 
\end{itemize}
{Leveraging} the above properties, {denoting by} $\tau_{GLRT-LAM}^{(\bar{N})}$ and $\tau_{P}^{(\bar{N})}$ the Left-Hand Side (LHS) and RHS of~\eqref{eq:CFAR_GLRT} computed assuming $P=\bar{N}$, respectively, {the following inequality holds}
\begin{equation}
\begin{aligned}
	\tau_{MFLRT-LAM} &= \max\limits_{1\le i \le \bar{N}} \left\{ g_i (l_G^{(i)}) \right\} \le  g_1 (l_G^{(\bar{N})}) \\
	&=g_1(2 (K+1) \tau_{GLRT-LAM}^{(\bar{N})}) \\ & \le g_1(2 (K+1) \tau_{P}^{(\bar{N})}) ,
\end{aligned}
\end{equation}
which {shows} that the detector~\eqref{eq_MFLRT} is bounded CFAR.

\subsection{{Two-Stage} Detectors/Estimators}
{Algorithms exploiting a linearization of the array manifold around the nominal search direction are well-know in open literature (see for instance~\cite{Nickel, AubryTSP}). Usually their performance depends on the distance between the true direction cosine value and that used for the {expansion}. For sufficiently high values of the mentioned displacement, a saturation is often experienced in the RMSE of the estimator when the Signal to Interference plus Noise Ratio (SINR) {is large enough}. To alleviate this phenomenon, a common (even if heuristic) approach relies on the use of a second stage {(also referred to as double stage)} of processing {based on} a re-linearization of the array manifold around the output of the first stage (single-{stage}) of processing (two-{stage} processing).	Generally, it yields some performance improvement{s} w.r.t. the single-iteration architecture.}
Therefore, for the case at hand, it is of practical interest to study the {capabilities} of the designed architectures when a further {linearization stage} is employed.
To enable the second {stage},} after the computation of the {angular} displacement estimate {$\widehat{\Delta u}_{\rm{LAM}}$} (as described in Section III)\footnote{{For the MFLRT-based procedure, $\widehat{\Delta u}_{\rm{LAM}}$ is the output of {\bf Algorithm~\ref{alg:1}} computed for {$P={\hat{i}}$, with ${\hat{i}}$} the {the index achieving the maximum in}~\eqref{eq_MFLRT}.}}, the ideal steering vector~\eqref{eq:standard_steering_vector} is {re-linearized around $\bar{u} + \widehat{\Delta u}_{\rm{LAM}}$}. 
Figs.~\ref{fig:flowchart} and~\ref{fig:flowchart_MFLRT} illustrate the flowchart of the procedures with reference to the GLRT and MFLRT detectors {(with and	without the second stage in red solid block and green dashed block, respectively)}.
Notice that the first iteration allows for the evaluation of $\tau_{GLRT-LAM}$ ($\tau_{MFLRT-LAM}$) and the immediate declaration of target presence/absence. {In addition}, as it will be illustrated in the numerical results, the detection and estimation tasks actually experience a performance boost thanks to the two-stage architecture.

\begin{figure*}[t]
	\centering
	\includegraphics[width=0.80\linewidth]{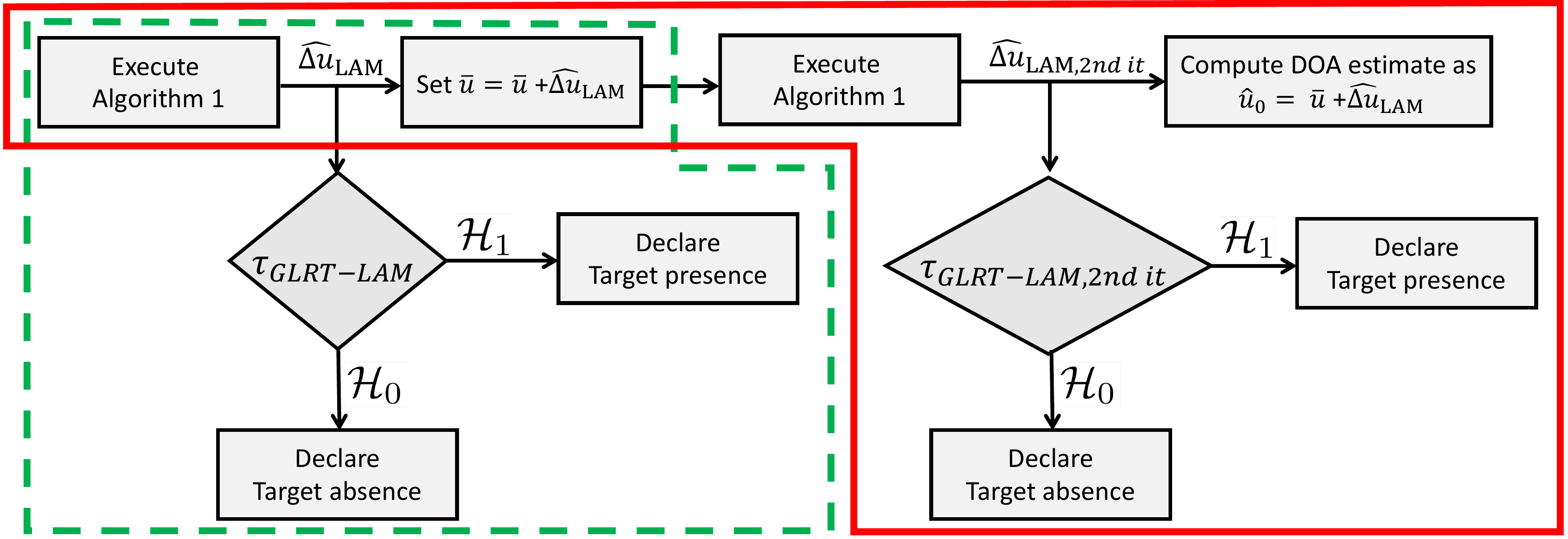}
	\caption{Flowchart of the GLRT-based procedures. The green dashed line demarcates the single iteration method, whereas the red solid one refers to the two-iterations processing.}\label{fig:flowchart}
\end{figure*}

\begin{figure}[t]
	\centering
	\includegraphics[width=0.90\linewidth]{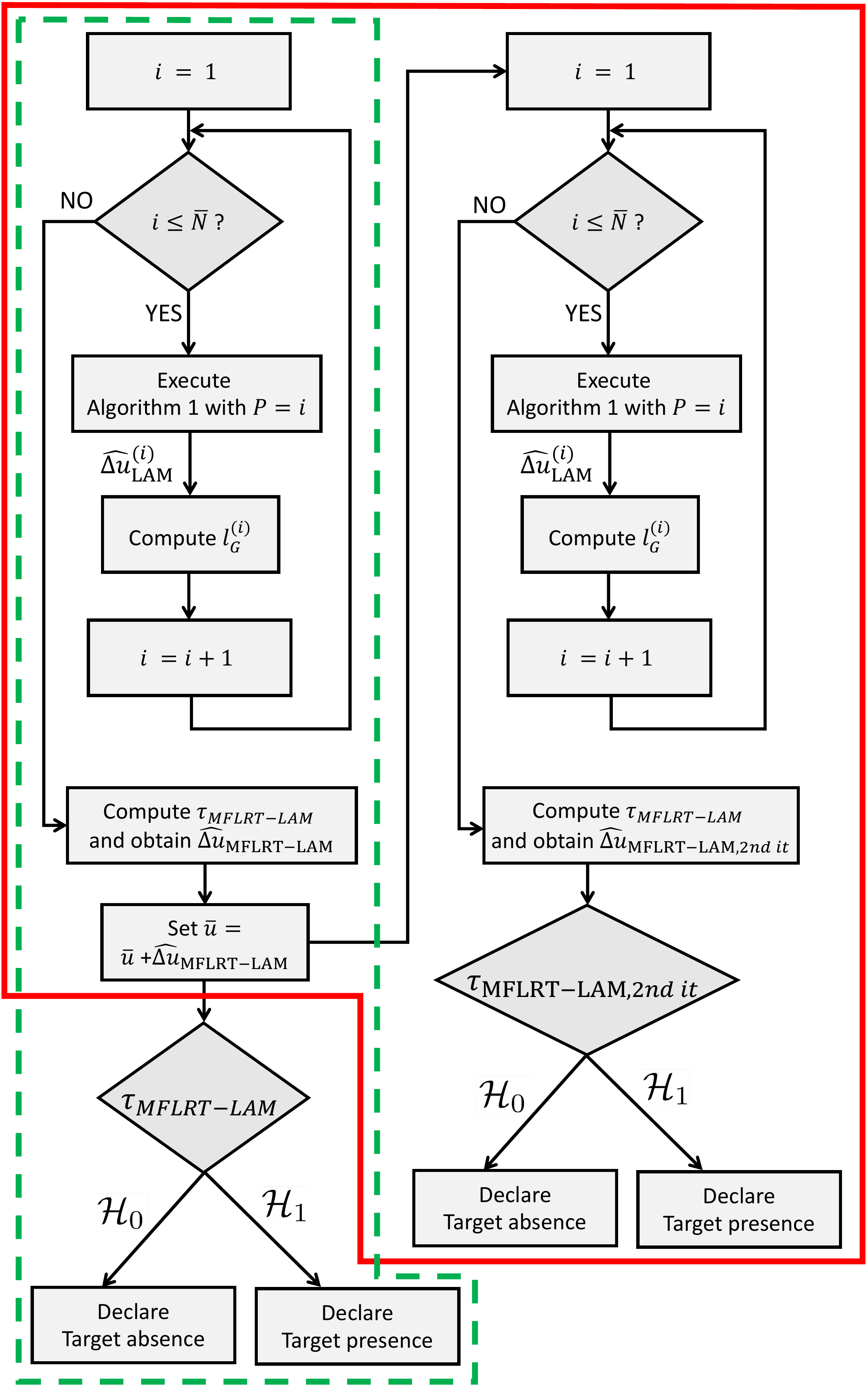}
	\caption{Flowchart of the MFLRT-based procedures. The green dashed line demarcates the single iteration method, whereas the red solid one refers to the two-iterations processing.}\label{fig:flowchart_MFLRT}
\end{figure}

\section{CRBs for ULA with mutual coupling}
In this section, the CRB for the unknown target DOA {displacement $\Delta u$} is derived, which is a key tool for the analysis of the statistical efficiency of the derived estimator {$\widehat{\Delta u}_{\rm{LAM}}$}. 
It is also worth mentioning that the CRB expression is {obtained} assuming known {interference} covariance matrix. However, considering the case of unknown $\bm{M}$, {which} is {a quantity functionally independent} on the target parameters, it will result in the same CRB expression for $\Delta u$ due to the block-diagonal structure of the corresponding Fisher Information Matrix (FIM).

In the following, the CRB is studied both for the actual {array manifold} case {(}which provides a performance benchmark to the estimation of $\Delta u=u_0-\bar{u}$, with $u_0$ the unknown to estimate{)} and for the linearized case {(}which {yields} a lower bound on the {displacement estimation performance} when the linearized model is employed{).}

\subsection{CRBs for the {Actual} Model}
Let us consider the {actual} signal model~\eqref{eq:s_v_mutual_c}
\begin{equation}
\begin{aligned}
		\bm{r} = & a\bm{p}_{m}{(u_0)} + {n} =  \bm{B}\bm{p}(u_0) + \bm{n} =  \breve{\bm{D}}(u_0)\bm{b} + \bm{n}
	\end{aligned}
\end{equation}
{where}
\begin{equation}
	\breve{\bm{D}}{(u_0)} = [\bm{p}(u_0), \bm{D}_1 \bm{p}(u_0), \dots, \bm{D}_{P-1} \bm{p}(u_0)] \in \mathbb{C}^{N \times P}.
\end{equation}

Denoting by $\bm{\theta} = [u_0, \bm{b}_R^\mathrm{T}, \bm{b}_I^\mathrm{T}]^\mathrm{T} \in \mathbb{R}^{2P+1}$ the vector of the real-valued unknowns, the FIM $\bm{F} \in \mathbb{R}^{(2P+1)\times (2P+1)}$ {can be} computed using the Slepian-Bangs formula \cite[p. 927, eq. 8.34]{vantrees4}, {as follows}
\begin{equation}\label{eq:Slepian_Bangs}
		\begin{split}
			{\bm{F}} &= 2\mathfrak{R} \left\{ {{{\left( {\frac{{\partial \breve{\bm{D}}{(u_0)}\bm{b}}}{{\partial {\bm{\theta }^{\mathrm{T}}}}}} \right)}^\dag }{{\bm{M}}^{ - 1}}\left( {\frac{{\partial   \breve{\bm{D}}{(u_0)}\bm{b}}}{{\partial {\bm{\theta }^{\mathrm{T}}}}}} \right)} \right\}\\
			&=2\mathfrak{R}\left\{  \left[ {\breve{\dot{\bm{D}}}}{(u_0)} \bm{b}, \breve{\bm{D}}{(u_0)}, j \breve{\bm{D}}{(u_0)} \right]^\dag  {{\bm{M}}^{ - 1}} \right .\\ & \left. \;\quad \qquad \left[ {\breve{\dot{\bm{D}}}{(u_0)}} \bm{b}, \breve{\bm{D}}{(u_0)}, j \breve{\bm{D}}{(u_0)} \right]     \right\},
		\end{split}
\end{equation}
where $\bm{b}_R = \mathfrak{R} \left\{ \bm{b}\right\}$, $\bm{b}_I = \mathfrak{I} \left\{\bm{b}\right\}$, and
${\breve{\dot{\bm{D}}}}{(u_0)} =  [\dot{\bm{p}}_u (u_0), \bm{D}_1 \dot{\bm{p}}_u (u_0), \dots, \bm{D}_{P-1} \dot{\bm{p}}_u (u_0)] \in \mathbb{C}^{N \times P}$ the derivative of $\breve{\bm{D}}$}.
Thus, the CRB for the target DOA is given by the first diagonal element of $\bm{F}^{-1}$ or alternatively, {after} partitioning $\bm{F}$ as
\begin{equation}
	\bm{F} = \begin{bmatrix}
		F_{uu} & \bm{F}_{ub} \\ 
		\bm{F}^\dagger_{ub}&\bm{F}_{bb}
	\end{bmatrix},
\end{equation}
it can be computed {as}~\cite{vantrees4}
\begin{equation}
	\textrm{CRB}(\Delta u) = \left[ F_{uu} - \bm{F}_{ub}\bm{F}_{bb}^{-1} \bm{F}^\dagger_{ub} \right]^{-1} ,
\end{equation}
where
\begin{equation}
F_{uu} = 2 \mathfrak{R} \left\{\bm{b}^\dagger {\breve{\dot{\bm{D}}}}^\dagger  {{\bm{M}}^{ - 1}} {\breve{\dot{\bm{D}}}} \bm{b} \right\},
\end{equation}
\begin{equation}
	\bm{F}_{ub} = 2\mathfrak{R}\left\{\bm{b}^\dagger {\breve{\dot{\bm{D}}}}^\dagger  {{\bm{M}}^{ - 1}} \left[ \breve{\bm{D}}, j \breve{\bm{D}} \right]     \right\},
\end{equation}
and
\begin{equation}
\bm{F}_{bb} = 2\mathfrak{R}\left\{  \left[\breve{\bm{D}}, j \breve{\bm{D}} \right]^\dag  {{\bm{M}}^{ - 1}} \left[ \breve{\bm{D}}, j \breve{\bm{D}} \right]     \right\}.
\end{equation}

\subsection{CRB for the Linearized Signal Model}
Assuming the useful target echo in the primary data modeled as {in}~\eqref{eq:B_pa} and invoking again the Slepian-Bangs formula \cite[p. 927, eq. 8.34]{vantrees4}, the CRB for the unknown DOA displacement $\Delta u$ is given by
\begin{equation}
	\textrm{CRB}_\textrm{LIN}(\Delta u) = \left[ F_{\Delta u \Delta u} - \bm{F}_{\Delta u b}\bm{F}_{bbLIN}^{-1} \bm{F}^\dagger_{\Delta ub} \right]^{-1},
\end{equation}
where
\begin{equation}
	F_{\Delta u \Delta u} = 2 \mathfrak{R} \left\{\bm{b}^\dagger {{\dot{\bm{D}}}}^\dagger  {{\bm{M}}^{ - 1}} {{\dot{\bm{D}}}} \bm{b} \right\},
\end{equation}
\begin{eqnarray}
	\bm{F}_{\Delta u b} = 2\mathfrak{R}\left\{\bm{b}^\dagger {{\dot{\bm{D}}}}^\dagger  {{\bm{M}}^{ - 1}} \left[ \tilde{\bm{D}} + \Delta u {\dot{\bm{D}}}, j (\tilde{\bm{D}} + \Delta u {\dot{\bm{D}}}) \right]     \right\},
\end{eqnarray}
and
\begin{equation}
\begin{aligned}
	\bm{F}_{bbLIN} = 2\mathfrak{R}&\left\{ \left[ \tilde{\bm{D}} + \Delta u {\dot{\bm{D}}}, j(\tilde{\bm{D}} + \Delta u {\dot{\bm{D}}})\right]^\dag  {{\bm{M}}^{ - 1}} \right. \\
	 & \; \left. \left[ \tilde{\bm{D}} + \Delta u {\dot{\bm{D}}}, j (\tilde{\bm{D}} + \Delta u {\dot{\bm{D}}}) \right]  \right\},
\end{aligned}
\end{equation}
{with $\tilde{\bm{D}}$ and $\dot{\bm{D}}$ defined as in~\eqref{eq:D_tilde} and~\eqref{eq:D_point}, respectively.}

\section{Performance Analysis}\label{section:perf_ana}
In this section, numerical examples are provided to evaluate both the detection and estimation capabilities of the devised processors {for} a ULA experiencing mutual coupling {among} its array elements. In the considered {experiments}, $N=16$, $K=3N=48$, and $\alpha= u_{3dB} \triangleq 0.891/N = 0.0557$. {The choice of $\alpha$ is a compromise between DOA uncertainty and quality of the linear approximation~\cite{AubryTSP}; although a specific value can be considered for each looking direction to account for the resulting Single-Side Beam Width (SSBW), a reasonable option could be considering the 3 dB SSBW $u_{3dB}$ of the ULA pointing at the boresight direction, regardless of the actual looking angle}. It is also assumed that the array pointing direction is set to $\theta = 35^\circ$, i.e., $\bar{u}=0.5736$, while the actual direction of the target is $u_{0}=0.6085$; therefore $\Delta u=0.0349$. 

Let us {model} the interference scenario assuming two narrow-band jammers located at $u_1=0.866$ and $u_2=-0.342$, respectively. As a consequence, the {interference-plus-noise} covariance matrix {is}
\begin{equation}
	{\bm{M}} = {\bm{\Sigma}_J} + \sigma_n^2{\bm{I}}_{N}
\end{equation}
with $\sigma_n^2$ the internal noise power level (assumed without loss of generality equal to 0 dB) and
\begin{equation}
	{\bm{\Sigma}}_J = \sum_{i=1}^2 \sigma_i^2 \bm{p}_m(u_{i}) \bm{p}_m^\dag(u_{i}),
\end{equation}
where $\sigma_1^2$ and $\sigma_2^2$ denotes the powers of {interferers}, with $\sigma_1^2/\sigma_n^2 = 30$ dB and $\sigma_2^2/\sigma_n^2 = 40$ dB, respectively, while $\bm{p}_m(u_{i})$ indicates the actual steering vector of the $i$-th ($i=1,2$) interfering source.

As to the mutual coupling, $P=3$ and the model coefficients are given by the vector ${[c_1, c_2]}^\mathrm{T} = [0.7, 0.4]^\mathrm{T}$.

Finally, the SINR is defined as
\begin{equation}
	\mathrm{SINR} = |a|^2\bm{p}_m^\dag(u_{0}){{\bm{M}}^{ - 1}}{\bm{p}_m}(u_{0}).
\end{equation}

The detection performance, reported in terms of $P_d$ versus SINR, is evaluated resorting to 1000 Monte Carlo (MC) runs, with $P_{fa}$ set to $10^{-4}$. In this regard, ${100}/{P_{fa}}$ MC trails are used to evaluate the detection thresholds. 
Furthermore, the angular estimation performance is assessed using the RMSE as figure of merit, computed as
\begin{equation}\label{eq:RMSE_MonteCarlo_u}
	\widehat{\text{RMSE}} = \sqrt{\frac{1}{\mathrm{MC}} \sum_{l=1}^{\mathrm{MC}} \left\|\Delta u-\widehat{\Delta u}_l\right\|^2} ,
\end{equation}
where $\widehat{\Delta u}_l$ is {the displacement} estimate at the $l$-th trial {and ${\mathrm{MC}} = 1000$}. In this context, for the estimation capability {of} the GLRT detector $\widehat{\Delta u}_l = \widehat{\Delta u}_{\rm{LAM}}$, {whereas} {$\widehat{\Delta u}_l = \widehat{\Delta u}_{\rm{LAM}}^{{\hat{i}}_l}$ {is considered} for the MFLRT processing with ${\hat{i}}_l$ the estimated model order {at the $l$-th trial}, i.e., the maximizer of~\eqref{eq_MFLRT}.} The MFLRT-based detector is {implemented} assuming four different values of $\bar{N}$, i.e., $\bar{N} \in \{2,4,6,8\}$.
 Moreover, the two-stage (referred to as ``2S'') version of both the GLRT and MFLRT is also considered. In the figures, the value of $\bar{N}$, employed for the execution of the MFLRT-based detectors, is specified as subscript.

{Finally,} for comparison purposes, the following detectors have been {contemplated}:
\begin{itemize}
	\item the GLRT using the actual array manifold with known target DOA and known coupling coefficients
	\begin{eqnarray}
		\tau_{\textrm{ben-GLRT}}= \frac{|\bm{r}^\dagger \bm{S}^{-1} \bm{p}_m({u}_0)|^2}{(1+ \bm{r}^\dagger \bm{S}^{-1} \bm{r}) \bm{p}_m^\dagger({u}_0) \bm{S}^{-1}\bm{p}_m({u}_0)} ;
	\end{eqnarray}
	\item the GLRT using the ideal array manifold (no coupling) with known target DOA
		\begin{eqnarray}
		\tau_{\textrm{ben-GLRT-NC}}= \frac{|\bm{r}^\dagger \bm{S}^{-1} \bm{p}({u}_0)|^2}{(1+ \bm{r}^\dagger \bm{S}^{-1} \bm{r}) \bm{p}^\dagger({u}_0) \bm{S}^{-1}\bm{p}({u}_0)};
	\end{eqnarray}
	\item {the GLRT using the actual array manifold with known target DOA and estimated coupling coefficients
	{\begin{equation}
		\tau_{\textrm{ben-GLRT-DOA}} = \frac{\bm{r}_w^\dagger \bm{{P}}_{\bar{D}_w} (u_0) \bm{r}_w}{1+\|\bm{r}\|^2},
	\end{equation}
where $\bm{{P}}_{\bar{D}_w} (u_0) = \bm{\bar{D}}_{w} \left( \bm{\bar{D}}_{w}^\dagger \bm{\bar{D}}_{w}\right)^{-1} \bm{\bar{D}}_{w}^\dagger$
with $\bm{\bar{D}}_{w} =  \bm{S}^{-1/2} [\bm{p}(u_0), \bm{D}_1 \bm{p}(u_0), \dots, \bm{D}_{P-1} \bm{p}(u_0)]$}};
	\item the standard GLRT using the ideal array manifold with the nominal pointing direction $\bar{u}$ (which {refers to a fully} mismatched case)~\cite{Kelly2}
	\begin{equation}
		\tau_{\textrm{GLRT}} = \frac{|\bm{r}^\dagger \bm{S}^{-1} \bm{p}|^2}{(1+ \bm{r}^\dagger \bm{S}^{-1} \bm{r}) \bm{p}^\dagger(\bar{u}) \bm{S}^{-1}\bm{p}};
	\end{equation}
	 \item the Subspace Detector (SD) \cite{Kraut}, {{namely} a GLRT detector which uses the ideal {linearized} array manifold (no coupling) and estimates the target displacement without imposing any constraints on $\Delta u$}
	 \begin{equation}
	\tau_{\textrm{SD}}=\frac{\bm{r}^\dagger \bm{S}^{-1} \bm{H}_{SD} \left(\bm{H}^\dagger_{SD} \bm{S}^{-1} \bm{H}_{SD} \right)^{-1} \bm{H}^\dagger_{SD} \bm{S}^{-1} \bm{r}}{1+ \bm{r}^\dagger \bm{S}^{-1} \bm{r}},
	 \end{equation}
\end{itemize}
with $\bm{H}_{SD} = [\bm{p}, \dot{\bm{p}}_u]$.

\subsection{Detection and Estimation Performance for Different number of Secondary Data}
\begin{figure*}[htbp]
	\centering
	\subfloat[]{\includegraphics[width=0.40\linewidth]{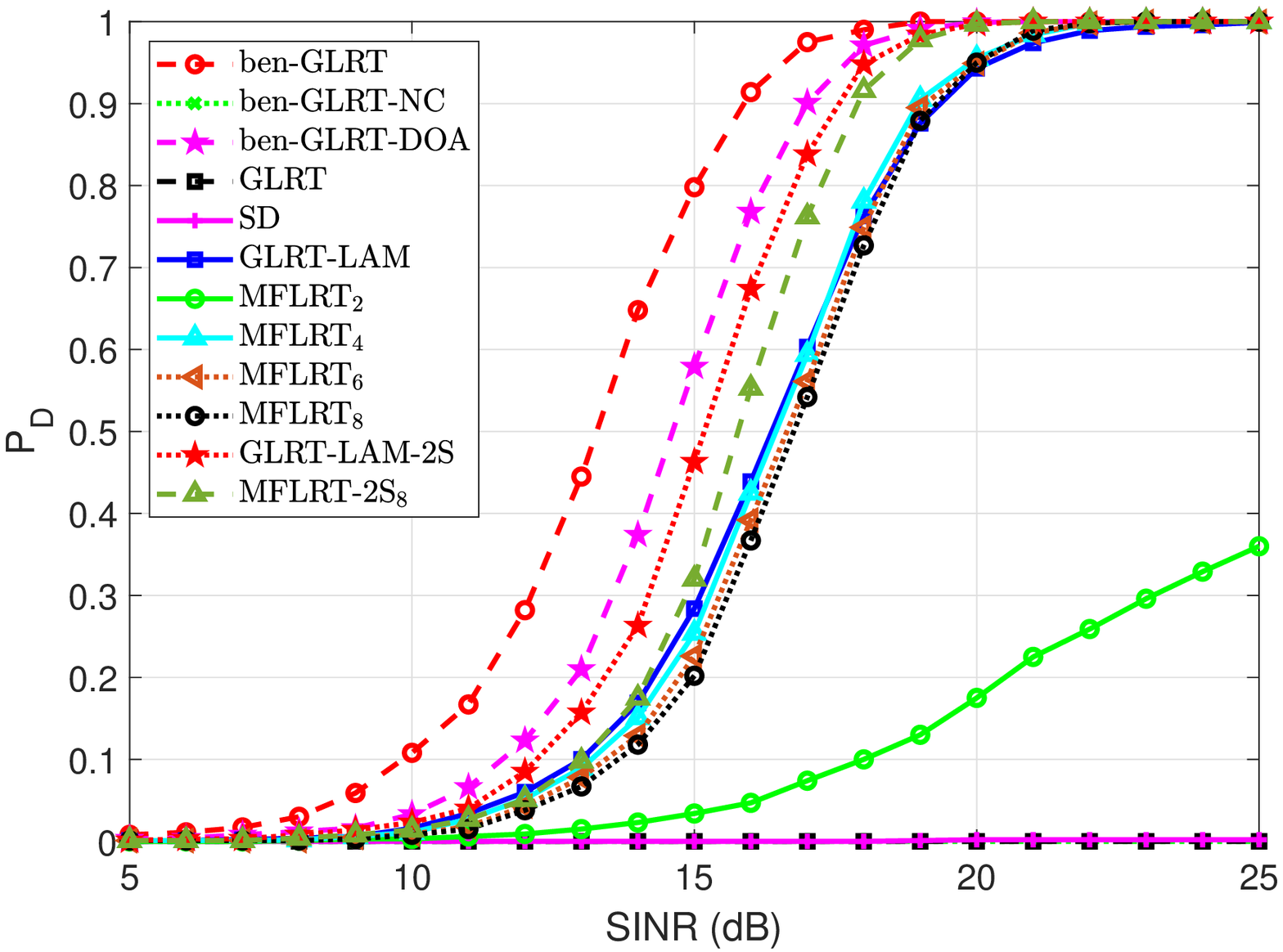} }
	\hfil
	\subfloat[]{\includegraphics[width=0.40\linewidth]{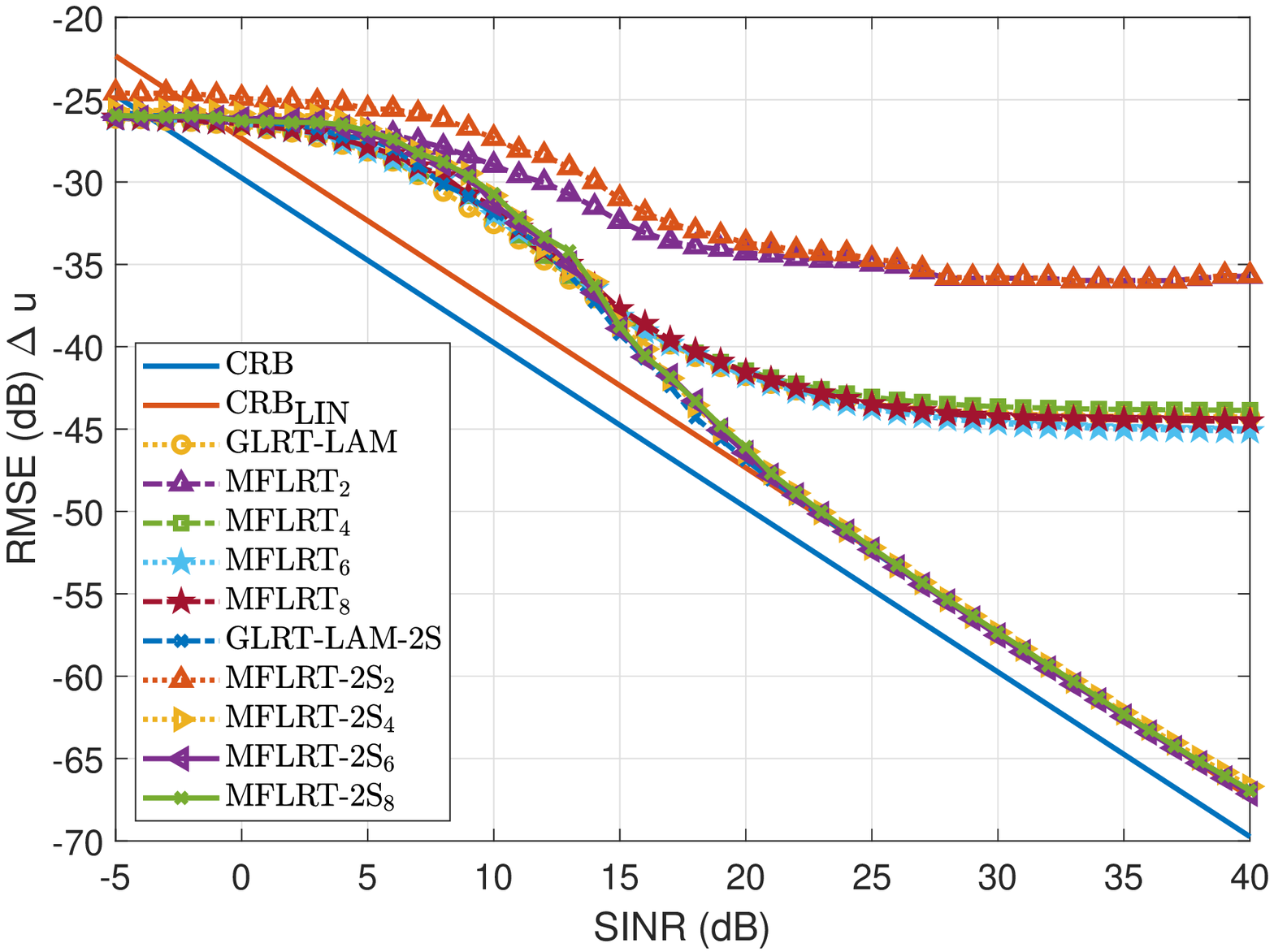} }
	\hfil \\
	\subfloat[]{\includegraphics[width=0.40\linewidth]{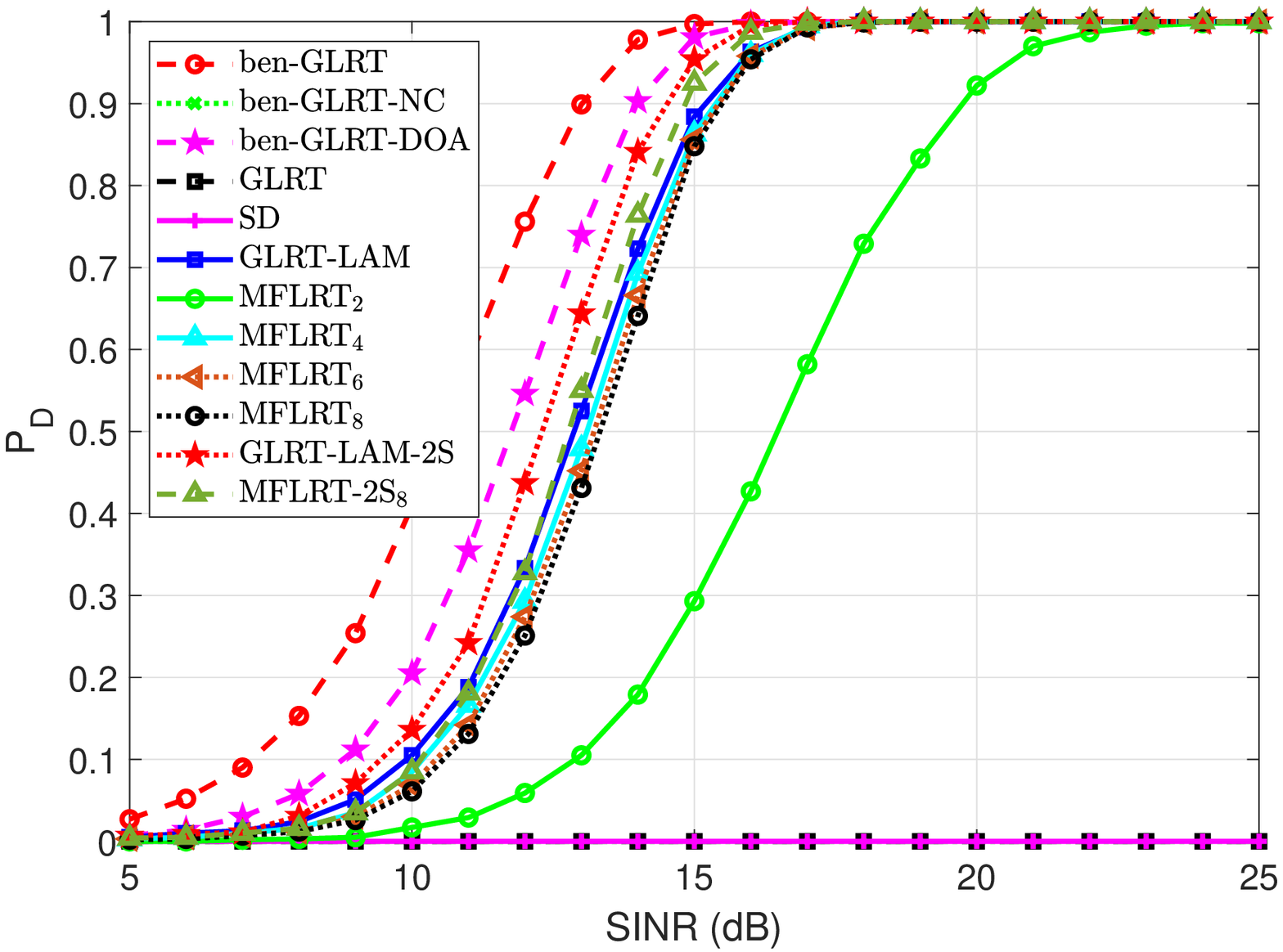} }
	\hfil
	\subfloat[]{\includegraphics[width=0.40\linewidth]{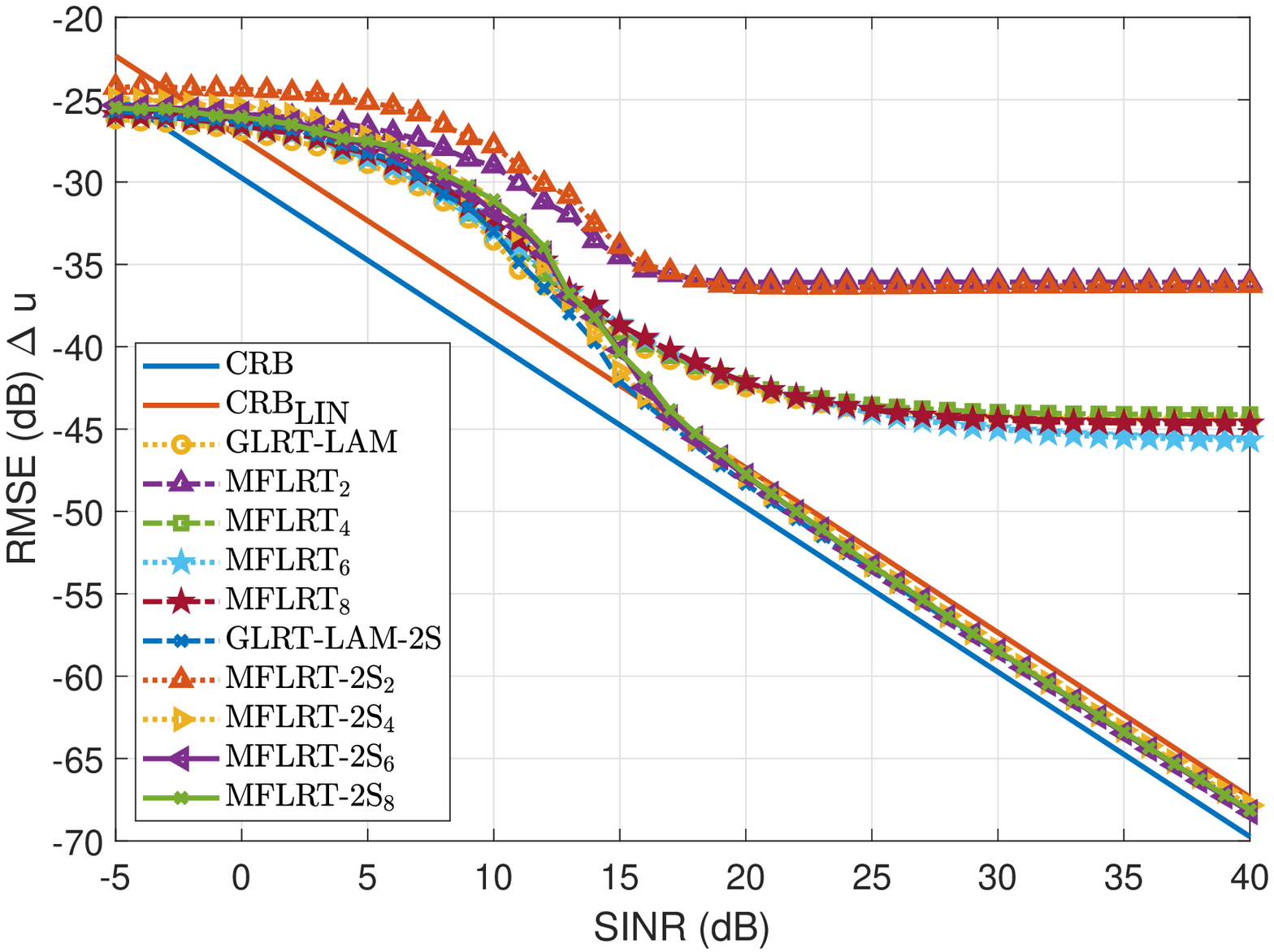} }
	\hfil
	\caption{Detection and estimation performance for a ULA with $N=16$ assuming $P=3$, $c_1=0.7$, and $c_2 = 0.4$. Figs. (a) and (c) report $P_d$ vs SINR while Figs. (b) and (d) depict RMSE (dB) vs SINR. Besides, Figs. (a) and (b) consider $K=32$, whereas (c) and (d) $K=80$.}
	\label{fig:pd_rmse_vs_sinr}
\end{figure*}

In Fig.~\ref{fig:pd_rmse_vs_sinr} the detection and estimation capabilities of the proposed signal processing architectures {are provided} in terms of $P_d$ and RMSE versus SINR.
Specifically, Fig{s}.~\ref{fig:pd_rmse_vs_sinr}(a) {and \ref{fig:pd_rmse_vs_sinr}(b)} consider $K=32$ secondary data, while Figs.~\ref{fig:pd_rmse_vs_sinr}(c) {and \ref{fig:pd_rmse_vs_sinr}(d)} refer to $K=80$.
Inspection of the $P_d$ curves reveals that the performance of the {single-stage GLRT-LAM and the MFLRT-based detectors is very close to each other (apart from the case of $\bar{N}=2$) with a loss, for $P_d=0.9$, of about 3 dB w.r.t. the ben-GLRT and in the order of 2 dB when compared with the ben-GLRT-DOA}. {This pinpoints} the capability of the devised methods to accomplish the detection task with satisfactory performance. {Additionally,} the results reveal the performance boost obtained by the two-stage versions of the GLRT-LAM and the MFLRT, with {a reduction}, in terms of SINR required to achieve $P_d=0.9$, {greater} than 1 dB w.r.t. the single-stage counterparts\footnote{For ease of visualization, in {Figs.}~\ref{fig:pd_rmse_vs_sinr} and~\ref{fig:pd_rmse_vs_delta} only the {$\textrm{MFLRT-2S}_8$} is displayed. However, the MFLRT-2S detectors with $\bar{N}=\{2,4,6\}$ {exhibit similar performance improvements w.r.t. their single stage counterparts as those resulting for the MFLRT with $\bar{N}=8$.}}. {Not surprisingly, for both single and double stage schemes, the devised GLRT-based detectors show a performance improvement w.r.t. the MFLRT counterparts, {due to} the capitalization of the prior knowledge on the model order}. 
Furthermore, in all the analyzed cases, the {receivers} {neglecting the} effect of mutual coupling, i.e., SD, GLRT and ben-GLRT-NC, are unable to provide adequate detection {capabilities} even {at} a high SINR regime, further stressing the need for tailored decision statistics that can {compensate for} the {unwanted} effect induced by mutual coupling.

Analysis of the estimation performance shows that the estimates provided by the devised single-stage methods deviate from the CRB for the linearized model and saturate in the high SINR regime. Remarkably, the two-stage versions of the GLRT-LAM and the MFLRT overcome {such a} shortcoming (by reducing the possible bias of the estimators) yielding RMSEs superimposed to the CRB for SINR $\ge$ 20 dB. 
Besides, the figures also show a gap (in the order of 2 dB) between the CRB curves for the actual and the linearized model, reflecting the presence of a signal modeling approximation.

It is also worth noting that the MFLRT with $\bar{N} = 2$ cannot provide a satisfactory detection performance due to its unavoidable underestimation of the model order which also causes a degradation in the estimation of the DOA displacement.

Finally, as expected, a comparison between Figs.~\ref{fig:pd_rmse_vs_sinr}(a) and \ref{fig:pd_rmse_vs_sinr}(c) {as well as} Figs.~\ref{fig:pd_rmse_vs_sinr}(b) and \ref{fig:pd_rmse_vs_sinr}(d), show that {increasing} the number of secondary data, the performance of all the reported procedures {improve}, due to the better estimate of the covariance matrix. {More specifically, {by} comparing the results for $K=80$ to those achieved for $K=32$, the detection performance improvement is {about} 3 dB for all the analyzed methods, while for the estimation task the gain is in the order of 1 dB.}

\subsection{Detection and Estimation Performance for Different $\Delta u$}
\begin{figure*}[htbp]
	\centering
	\subfloat[]{\includegraphics[width=0.40\linewidth]{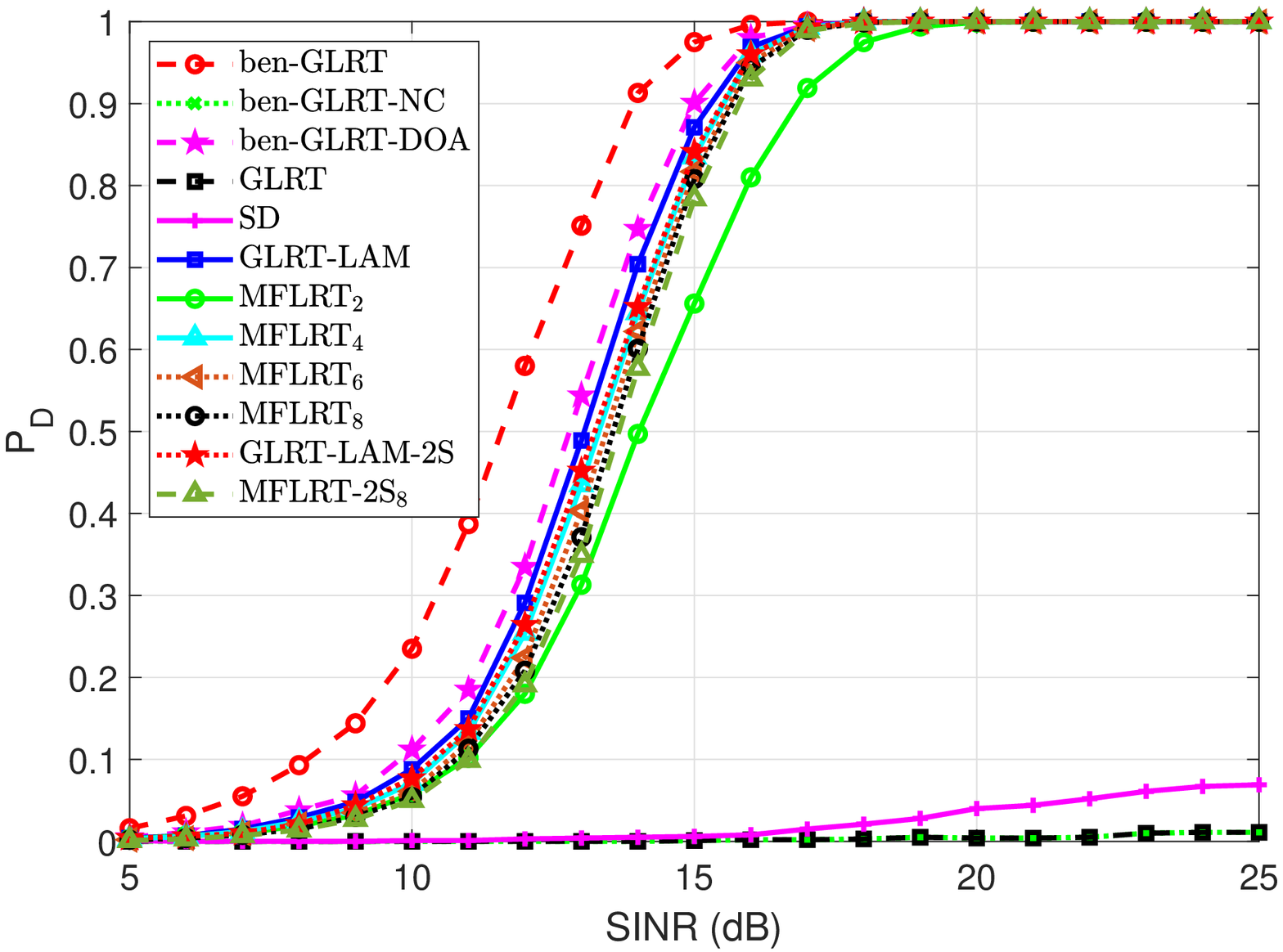} }
	\hfil
	\subfloat[]{\includegraphics[width=0.40\linewidth]{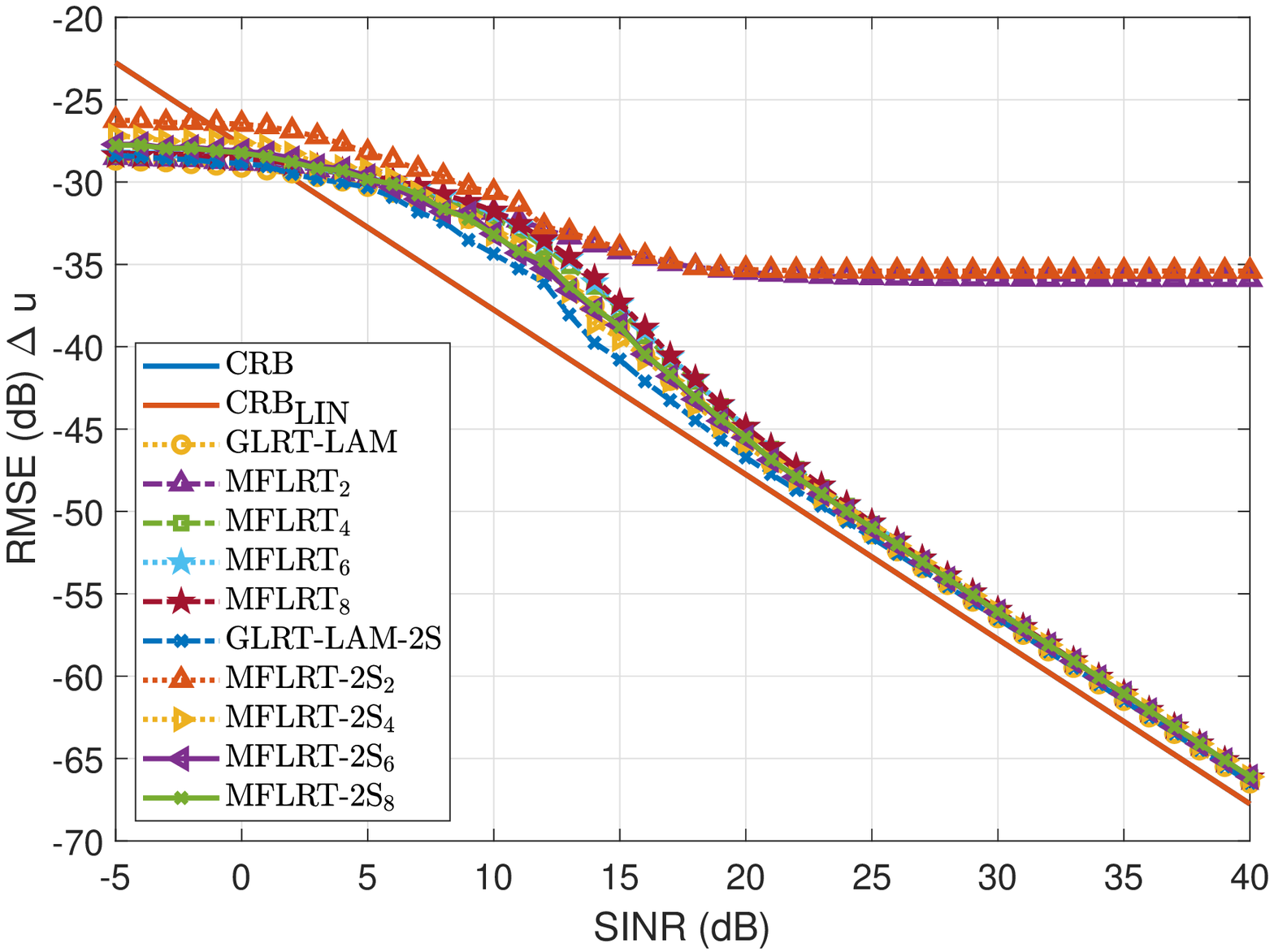} }
	\hfil \\
	\subfloat[]{\includegraphics[width=0.40\linewidth]{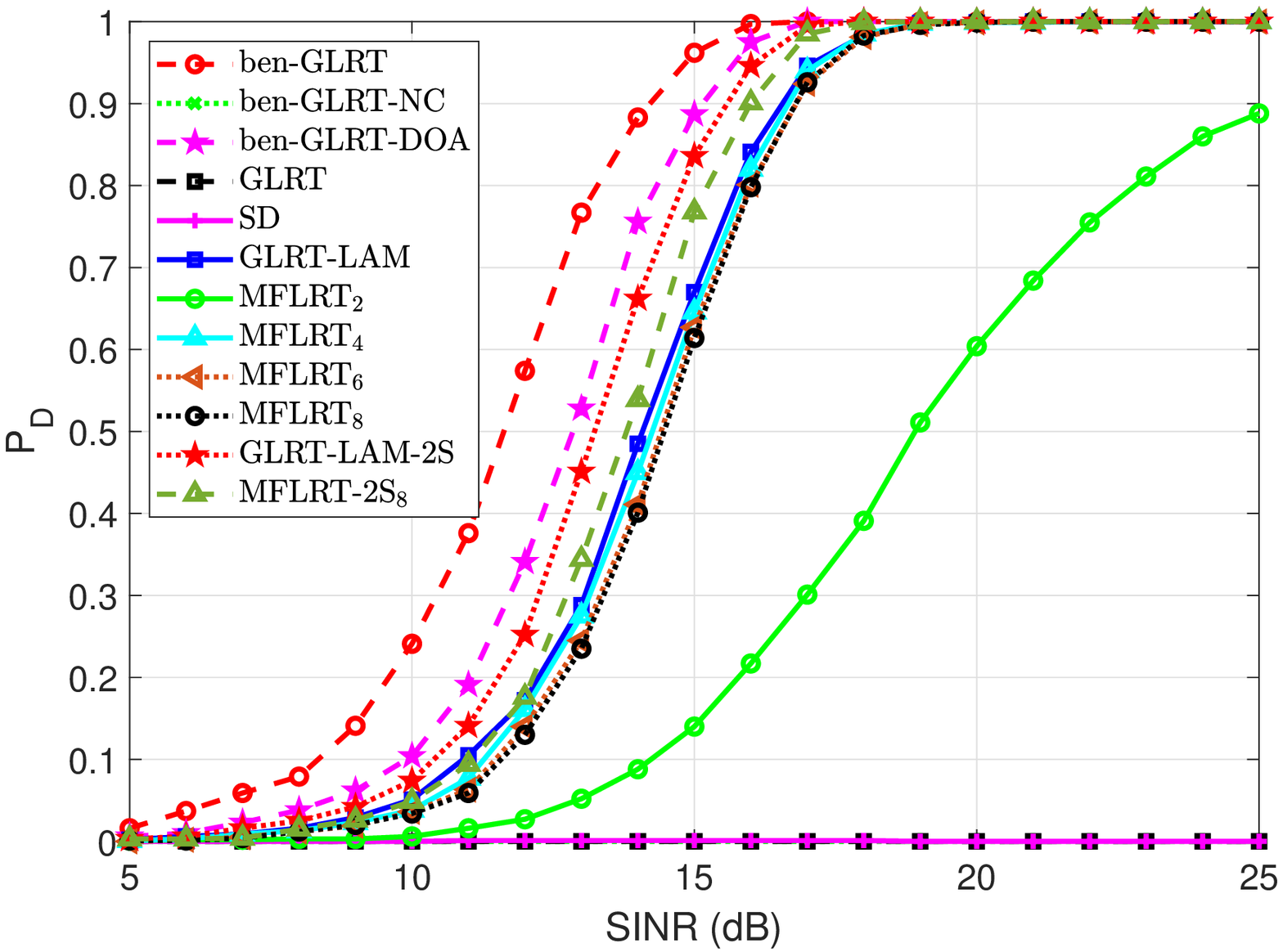} }
	\hfil
	\subfloat[]{\includegraphics[width=0.40\linewidth]{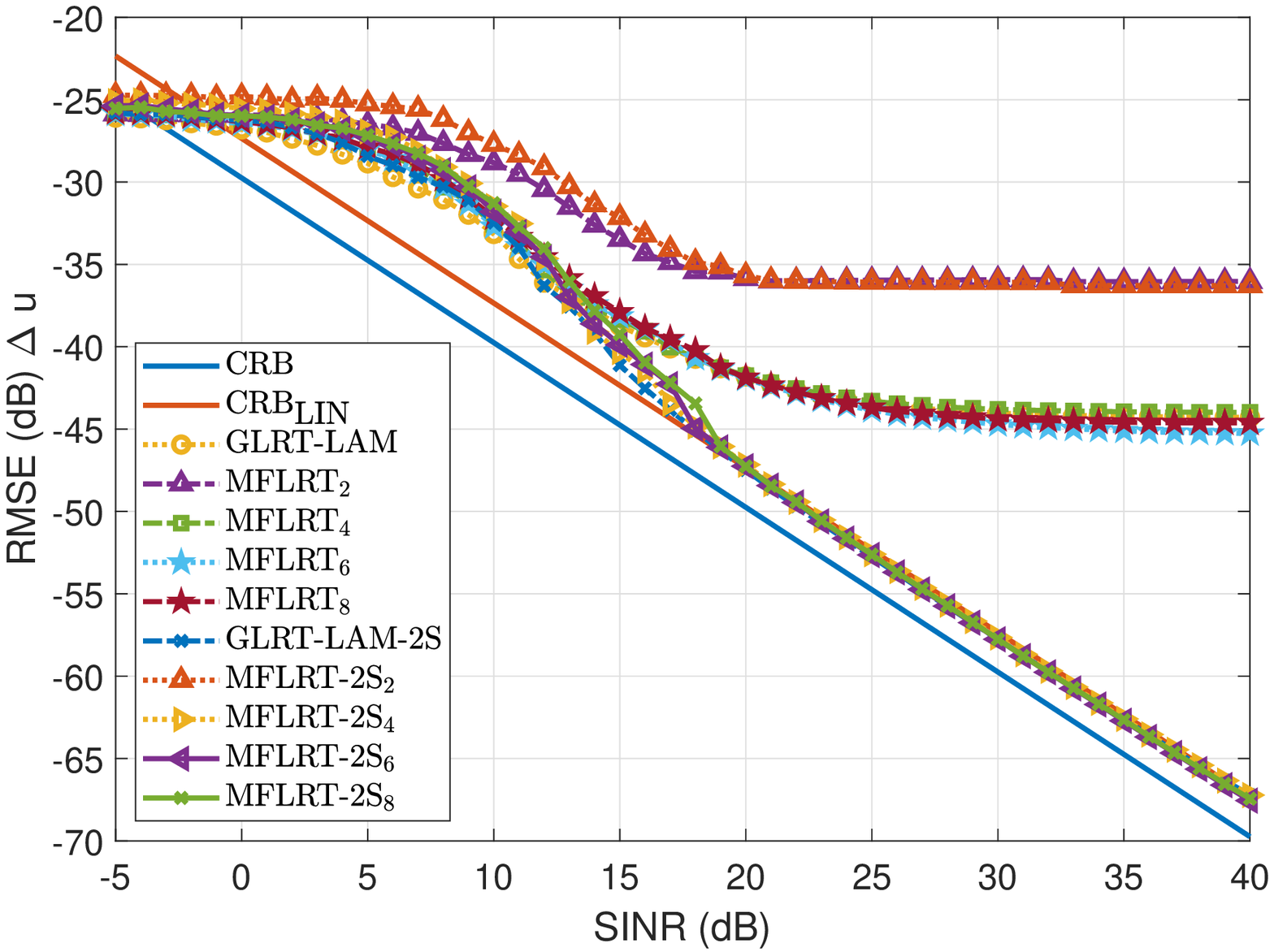} }
	\hfil
	\caption{Detection and estimation performance for a ULA with $N=16$ assuming $P=3$, $c_1=0.7$, $c_2 = 0.4$, $K=48$, and Figs. (a) and (b) $\Delta u = 0$, Figs. (c) and (d) $\Delta u = 0.0349$. Moreover, Figs. (a) and (c) report $P_d$ vs SINR while Figs. (b) and (d) illustrate RMSE (dB) vs SINR.}
	\label{fig:pd_rmse_vs_delta}
\end{figure*}

Fig.~\ref{fig:pd_rmse_vs_delta} reports the detection and estimation performance for the same scenario {as in} Fig.~\ref{fig:pd_rmse_vs_sinr} but assuming $K=48$ and two different values for the displacement, i.e., $\Delta u = 0$ in Figs.~\ref{fig:pd_rmse_vs_delta}(a) and \ref{fig:pd_rmse_vs_delta}(b), $\Delta u = 0.0349$ in Figs.~\ref{fig:pd_rmse_vs_delta}(c) and \ref{fig:pd_rmse_vs_delta}(d).
The results highlight that for the case of $\Delta u = 0$ (Figs.~\ref{fig:pd_rmse_vs_delta}(a) and \ref{fig:pd_rmse_vs_delta}(b)), the $P_d$ curves pertaining to the GLRT-LAM and the MFLRT with $\bar{N} \in \{4,6,8\}$ are almost superimposed to the ben-GLRT-DOA and are {quite} close to the ben-GLRT {performance}, {showing} a loss in the order of 1 dB {at} $P_d = 0.9$. For this case study, {the two-stage detectors {experience} a slight performance loss, which is totally in line with the rationale leading to the design of the two-stage architectures, since for this case study the best possible point {of the expansion} is already used at the first stage.}

{The effectiveness} of the methods {is} also {corroborated} by the RMSE versus SINR curves, reported in Fig.~\ref{fig:pd_rmse_vs_delta}(b), {which show an {estimation performance} close to the CRB (but {for} the MFLRT {approaches {with}} $\bar{N} = 2$)}. Moreover, for the considered case {study}, the CRB computed for the actual signal model is overlapped with that {obtained} for the linearized one. Interestingly, the MFLRT with $\bar{N}=2$ can still provide adequate detection performance, with a {SINR loss smaller} than 2 dB {as} compared to the GLRT-LAM.
The case of $\Delta u = 0.0349$ is illustrated in Figs.~\ref{fig:pd_rmse_vs_delta}(c) and \ref{fig:pd_rmse_vs_delta}(d), which {highlight} a detection and estimation performance similar to that in Fig.~\ref{fig:pd_rmse_vs_sinr}, {which refers to} a different sample support size $K$, i.e., $K=\{32, 80\}$. {Again, in the analyzed {scenarios}, the SD, the GLRT, and the ben-GLRT-NC detectors show underwhelming detection performance due to the {disregard} of the mutual coupling {phenomenon} at the design stage.}

\subsection{Cosine Similarity between the Actual and Estimated Steering Vectors}
\begin{figure}[htbp]
	\centering
	\subfloat[]{\includegraphics[width=0.78\linewidth]{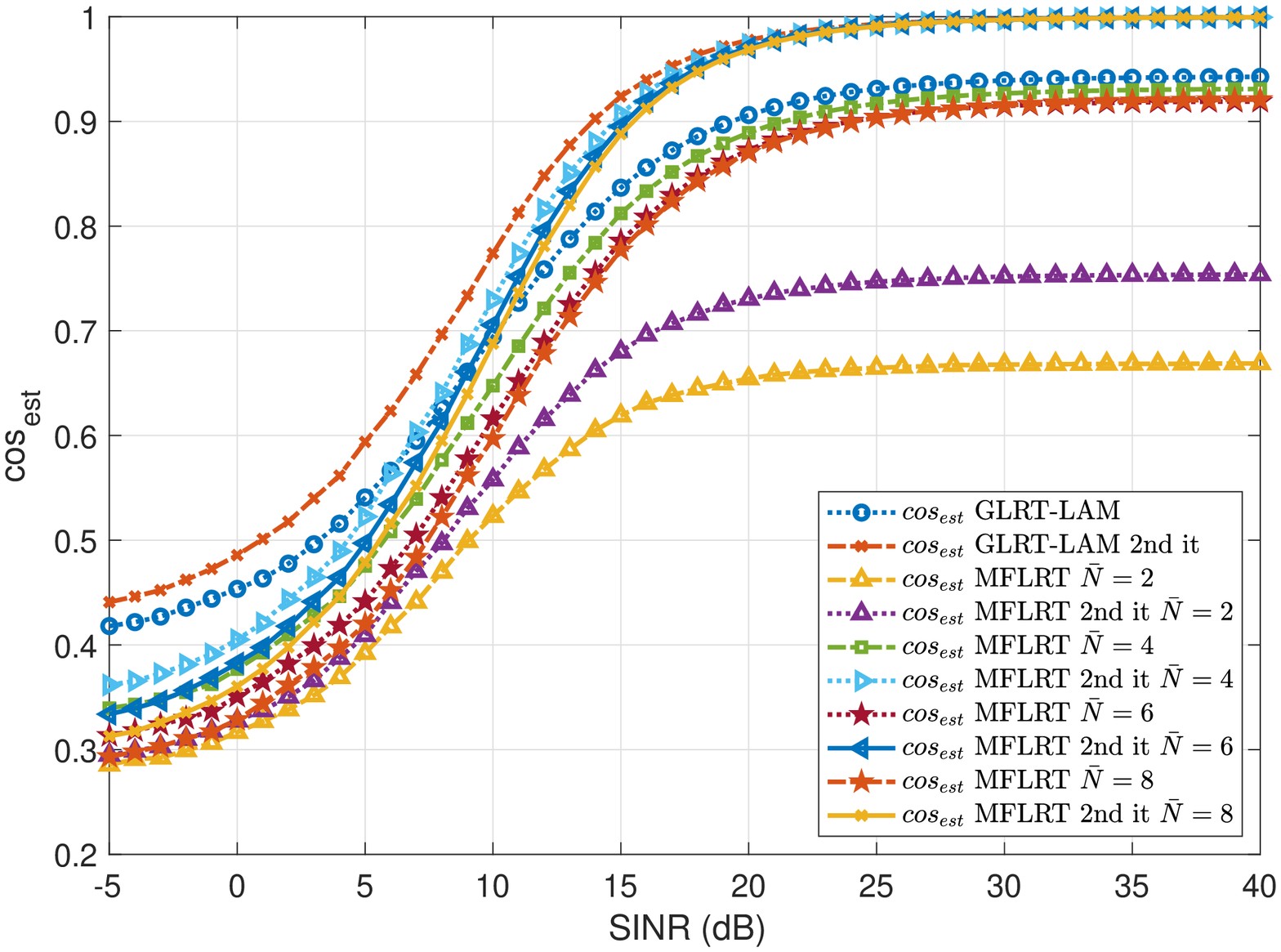} }
	\hfil	\\
	\subfloat[]{\includegraphics[width=0.78\linewidth]{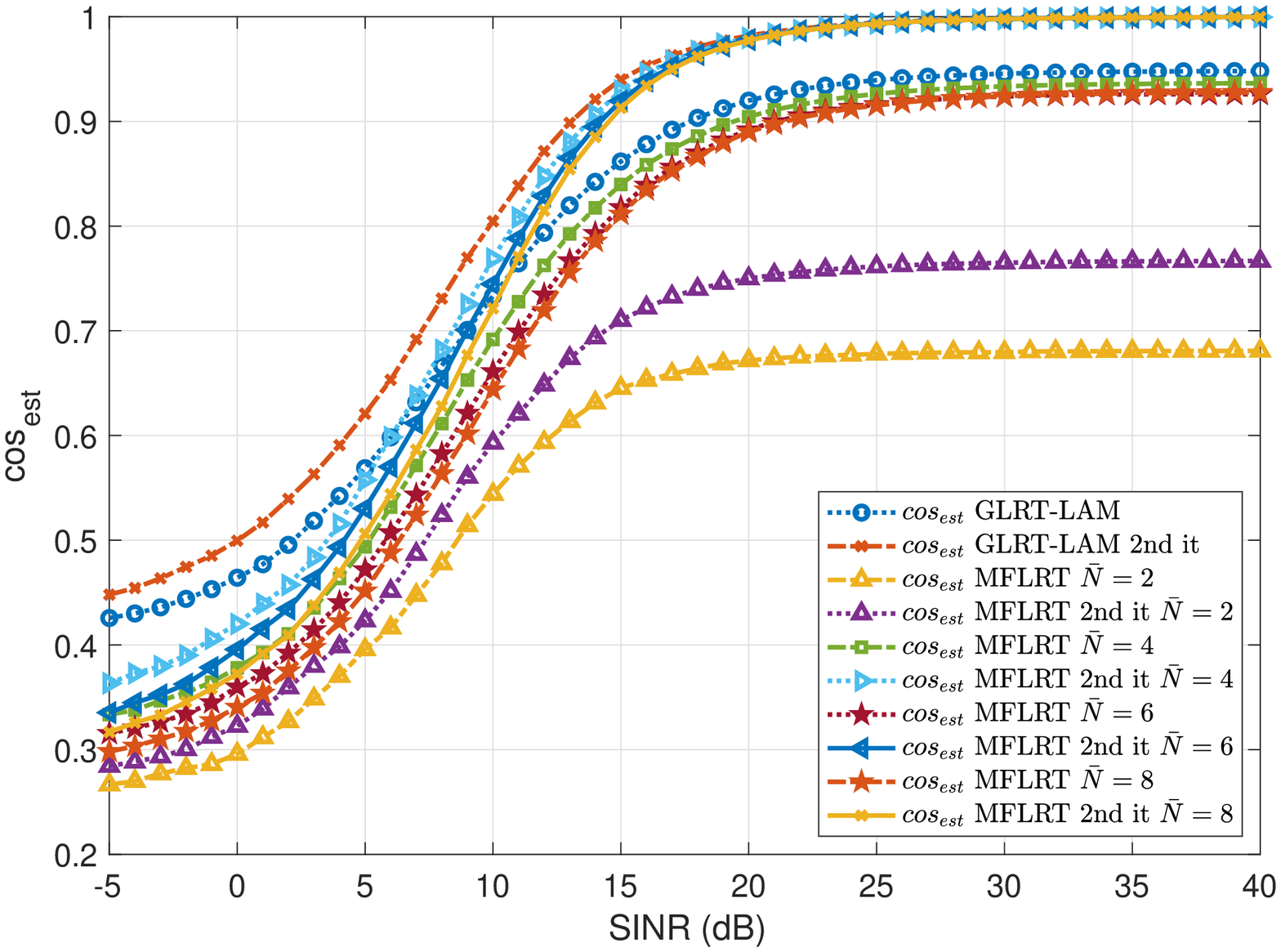} }
	\hfil \\
	\subfloat[]{\includegraphics[width=0.78\linewidth]{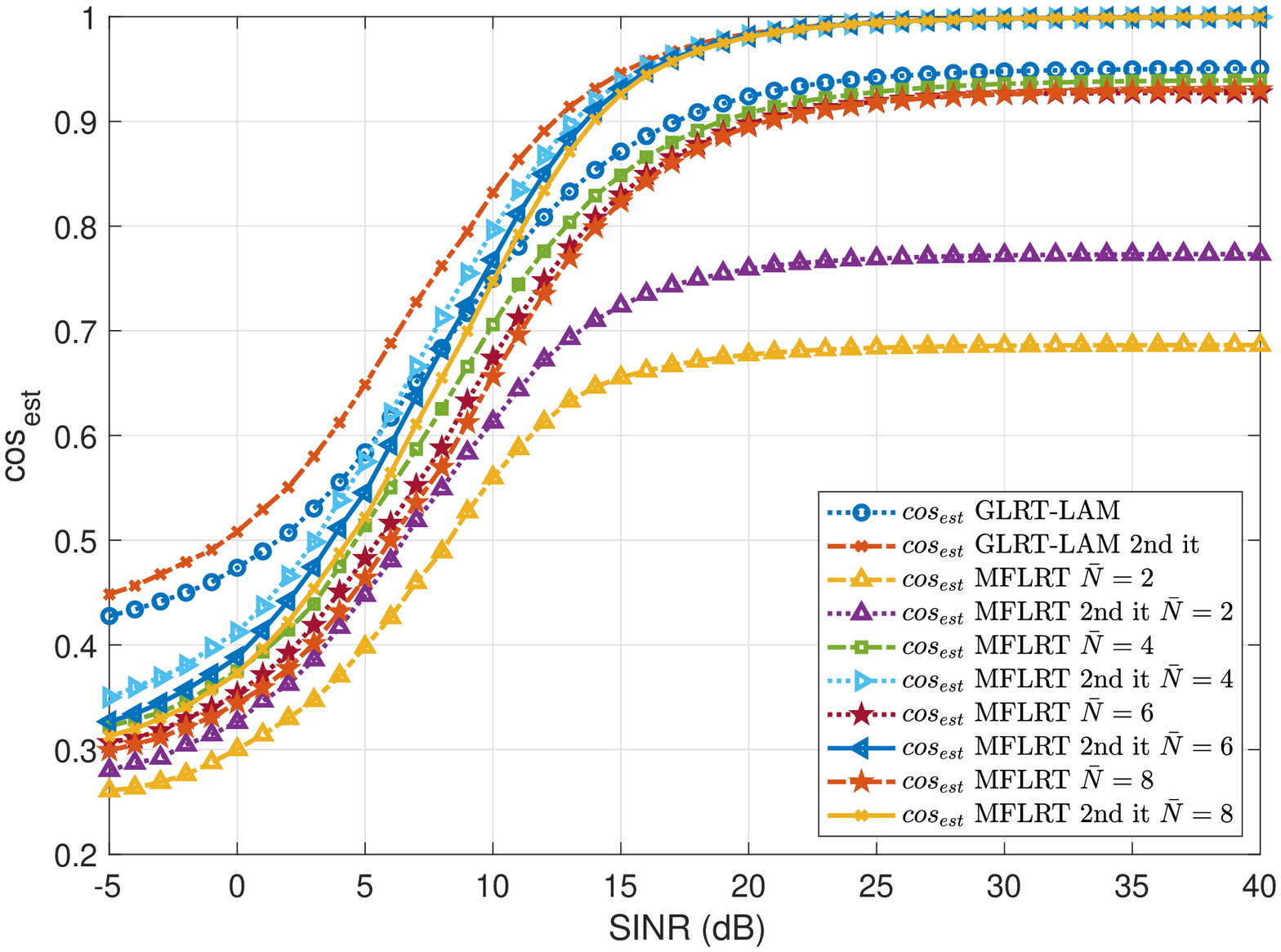} }
	\hfil
	\caption{Average cosine similarity {in the whitened signal space} between the actual steering vector and the estimated one for (a) $K=32$, (b) $K=48$, (c) $K=80$. }\label{fig:cos_s_vs_sinr}
\end{figure}

To further {assess} the {estimation} capabilities of the devised architectures, assuming the same configuration {as in} Fig.~\ref{fig:pd_rmse_vs_sinr}, Fig.~\ref{fig:cos_s_vs_sinr} reports, for $K \in \{32, 48, 80\}$, the {average} cosine similarity {in the whitened signal space} versus SINR between the actual steering vector and the one computed using the estimates of both the angular mismatch and the coupling coefficients involved in the evaluation of the GLRT-LAM and the MFLRT-based detectors. Specifically, {for a given SINR,} the {average} cosine similarity is {evaluated over 1000 MC trials} as
\begin{equation}
	\mathrm{cos}_{est} = {\frac{1}{MC} \sum_{l=1}^{MC}} \frac{{| \bm{p}_m(u_0)^\dagger \bm{M}^{-1} \hat{\bm{p}}_m(\hat{u}_l) |} }
{ \|\bm{M}^{-1/2} \bm{p}_m(u_0)\|     \|\bm{M}^{-1/2} \hat{\bm{p}}_m(\hat{u}_l)\|   }	,
\end{equation}
where, {at the $l$-th trial, }$\hat{\bm{p}}_m(\hat{u}_l) = \bm{\hat{C}}_l\bm{p}(\hat{u}_l)$ is the estimated steering vector with $\hat{u}_l={\bar{u}+\widehat{\Delta u}_l}$ and $\bm{\hat{C}}_l$ the estimate of the coupling matrix using the coupling coefficient vector $\bm{\hat{c}}_l = \hat{\bm{b}}_l/\hat{\bm{b}}_l(1)$.
The {developed analysis shows} that, regardless of the number of secondary data, {in the} high SINR regime, all the devised methods, with the exception of the MFLRT{s} with $\bar{N}=2$, are able to provide an adequate estimation of the steering vector, with values of {the} cosine similarity greater then 0.9. Notably, the two-stage version of {each} procedure {leads to} cosine similarity values {close to} 0.99, which {corroborates} the ability of the devised methods to {perform an accurate} estimate {of} both {the} DOA {displacement} and {the} mutual coupling coefficients. Finally, it is not surprising that, for a given SINR, as the secondary data increases, the covariance matrix estimates {become more reliable} leading to higher levels of {the} achieved cosine similarity (e.g., see Table~\ref{table:cos_sim}). 

\begin{table}[thp] 
	\centering
	\caption{Average cosine similarity in the whitened signal space between the actual and estimated steering vector for $\text{SINR}=15$ dB.}\label{table:cos_sim}
	\begin{tabular}{lccc}
		\hline
		\hline
		\textbf{Method} & \textbf{$K=32$} & \textbf{$K=48$}   & \textbf{$K=80$} \\
		\hline
GLRT-LAM          & 0.84 & 0.86 & 0.87 \\
MFLRT $\bar{N}=2$       & 0.62 & 0.65 & 0.66 \\
MFLRT $\bar{N}=4$        & 0.81 & 0.84  & 0.85 \\
MFLRT $\bar{N}=6$       & 0.79 & 0.82 & 0.83  \\
MFLRT $\bar{N}=8$        & 0.78 & 0.81 & 0.82 \\
GLRT-LAM 2S   & 0.92 & 0.94  & 0.95 \\
MFLRT $\bar{N}=8$ 2S & 0.89 & 0.91 & 0.93\\
		\hline
		\hline
	\end{tabular}
\end{table}

\section{Conclusions}
Assuming the presence of mutual coupling among the array elements, {joint} adaptive detection and DOA estimation of a prospective radar target {have} been considered. To this end, a bespoke model of the received signal has been developed, leveraging array manifold linearization around the nominal {look} direction as well as the description of the mutual coupling effects via symmetric Toeplitz matrices. As to the latter aspect, two situations have been considered so as to account for different amounts of a-priori information available on the mutual coupling {phenomenon}. The former assumes known the model order whereas the latter refers to the case where the coupling depth is unknown. 

Hence, appropriate adaptive architectures to detect targets and estimate the corresponding DOA have been designed for each situation. Specifically, resorting to advanced optimization tools, the GLRT detector has been synthesized when the model order is known while the MFLRT {is used} when the aforementioned information is not available. Notably, both the strategies exhibit a bounded-CFAR behavior.

Some interesting case studies have been illustrated to assess the capabilities of the novel devised architectures also in comparison with clairvoyant benchmarks as well as with detectors that do not model the presence of mutual coupling during their design process. Both detection probability and RMSE on the target bearing have been assessed, clearly highlighting the performance benefits offered by the synthesized mutual coupling robust receivers.

Possible future research avenues might be focused on the extension of the framework to two-dimensional arrays~\cite{Wu-coupling} as well as the analysis of the developed detectors in the presence of real and/or synthetic data obtained with a high-fidelity electromagnetic simulator accounting for mutual coupling.

\small

\input{./biblio.tex}

\end{document}

%% file: biblio.tex
\newpage{\pagestyle{empty}\cleardoublepage}